\documentclass[aps,pre,showpacs,floats,twocolumn,floats,superscriptaddress,floatfix,english]{revtex4-1}
\usepackage[T1]{fontenc}
\usepackage[latin9]{inputenc}
\usepackage{array}
\usepackage{multirow}
\usepackage{setspace}

\usepackage{graphicx}
\usepackage{subfigure}
\usepackage{color}
\usepackage{amsmath}
\usepackage{verbatim}
\usepackage{breakcites}
\usepackage{pgfplots}
\usepackage{natbib}
\usepackage{nameref}
\usepackage{xcolor}
\usepackage{hyperref}
\usepackage{cleveref}

\makeatletter

\providecommand{\tabularnewline}{\\}
\usepackage{caption}
\usepackage{color}
\usepackage{epstopdf}
\usepackage{epsfig}
\usepackage[normalem]{ulem}
\usepackage{babel}

\def\tb{\textcolor{black}}

\renewcommand{\sout}[1]{\unskip}

\newcommand{\nordita}{Nordita, KTH Royal Institute of Technology and Stockholm University, Stockholm 10691, Sweden}
\begin{document}
\title{Transient Convective Spin-up Dynamics}

\author{S. Ravichandran}
\affiliation{\nordita}
\author{J. S. Wettlaufer}
\affiliation{\nordita}
\affiliation{Yale University, New Haven, Connecticut 06520-8109, USA}
\makeatother
\begin{abstract}
We study the formation, longevity and breakdown of convective
rings during impulsive spin-up in square and cylindrical containers using direct numerical simulations.
The rings, which are axisymmetric alternating regions of up- and downwelling 
flow that can last for $\mathcal{O}\left(100\right)$ rotation times,
were first demonstrated experimentally and arise due to a balance between
Coriolis and viscous effects. We study the formation of these rings in the context
of the Greenspan-Howard spin-up process, the disruption of which  
modifies ring formation and evolution.   We show that, unless imprinted by boundary geometry, 
convective rings can only form when the surface providing
buoyancy forcing is a free-slip surface, thereby explaining an apparent disagreement between experimental
results in the literature. For Prandtl numbers from 1--5 we find that the longest-lived rings occur for 
intermediate Prandtl numbers, with a Rossby number dependence.  
Finally, we find that the constant evaporative heat-flux conditions imposed in the experiments are 
essential in sustaining the rings and in maintaining the vortices that form in consequence of the ring breakdown.
\end{abstract}
\maketitle

\section{Introduction \label{sec:Intro}}

The dynamical processes by which a fluid within a spinning container attains the same angular velocity as the vessel is referred to as the ``spin-up'' (or ``spin-down'') problem, and was unified in the theoretical treatment of \cite{greenspan_howard} (hereafter GH). Suppose that the vessel is a right solid of horizontal dimension $L$ containing an isothermal fluid of viscosity $\nu$.   At $t=0$ the container is rotated about its vertical axis with a constant angular velocity $\Omega$. The fluid takes a finite amount of time to 
\textquotedblleft spin up\textquotedblright{} to the angular velocity of the solid container. Clearly, were the required transfer of angular momentum controlled solely by viscosity the spin-up time would scale as ${\tau_s}^\nu \propto L^2 / \nu$.  However, GH showed that $\tau_s = \Omega^{-1} Re^{1/2}$, where the Reynolds number is $Re=L^2\Omega/\nu$, and hence ${{\tau_s}^\nu}/{\tau_s} \propto Re^{1/2}$.  Therefore, given that $Re$ is typically large, the time required for fluid spin-up is much smaller than if the process were controlled by viscosity alone. 
%
%\tb{GH showed that the spin-up time $\tau_s$ is in fact $\tau_s \propto L/\left(\Omega \nu \right)^{1/2} = \Omega^{-1} Re^{1/2}$. Thus, in units of $\Omega^{-1}$, the spin-up time is proportional not to the Reynolds number $Re=L^2\Omega/\nu$ but to the square root of the Reynolds number. (Note that if there is no temperature in the problem, the only nondimensional number parameterising the dynamics is the Reynolds number.) Since this is typically a large number, the time required for fluid spin-up is much smaller than if spin-up occurred due to viscosity alone. In the final state, the whole mass of fluid spins at the angular velocity $\Omega$.}\\

When the surfaces of the container are heated, the interplay between buoyancy and rotational forces complicates 
the dynamics considerably. For example, when the container is heated from below, the long-term ($\tau \gg \tau_s$) state is characterized by columnar vortices aligned in the direction of gravity, along which fluid is transported. Here, we study the spin-up of a convectively unstable impulsively rotated container of
fluid to its final vortical state. In particular, we are interested in a transient ring pattern that occurs during convective spin-up. This ringed state consists
of alternating axisymmetric rings of up- and downwelling flow, which have been reported in experimentally 
by \cite{boubnov1986}, \cite{vorobieff1998}, and \cite{ZPW}. 

The experiments  of \cite{boubnov1986} were performed in square and circular cross-sectioned containers of water with open upper surfaces cooled by evaporation. 
They measured the temperature of the free surface and estimated the rate of evaporation, and hence the cooling rate, to be nearly steady.  
When the upper surface was one of free slip, they observed the transient ringed state for a wide range of rotation and cooling rates (varied by changing the mean temperature of the water).  However, for both square and circular cross-sections, when the top surface was covered by a lid, and the bottom surface is heated, they found no ringed state, 

In contrast to  \cite{boubnov1986}, \cite{vorobieff1998} held the bottom surface at constant temperature and found
the ringed state (albeit with fewer rings) in a cylindrical container with a no-slip upper surface.
\cite{ZPW} combined particle image velocimetry with infrared thermometry in a square cross-section
container of depth $H$ with an evaporating free-slip upper surface. They quantified the ringed state as a transient
balance between rotational and viscous forces that exists for approximately one Ekman time, $\tau_E = \sqrt{H^2/\Omega \nu}$.

Here, we study the formation and breakdown of these transient convective rings using numerical simulations in a variety of geometries.  We find that the ringed state is a universal feature of convective spin-up, \tb{and that for certain boundary conditions, the rings take on the shape of the container, leading to square `sheets' of convection for square cross-sectioned geometries,}  and the Prandtl number plays an important role in the formation and stability of the rings. Additionally, we find that the thermal boundary conditions used--Dirichlet as in \cite{vorobieff1998} and Neumann as in \cite{boubnov1986} and \cite{ZPW}--influence the ring stability and the dynamics of their breakdown.  Our results reconcile the seemingly contradictory observations of \cite{boubnov1986} and \cite{vorobieff1998}. 

In \S \ref{sec:setup}, we describe the setup of the numerical simulations and the key differences from the experiments. We then discuss the numerical methods used and the
resolution requirements for the simulations. The formation, longevity and breakdown of the ringed state into the final vortical state is summarized in \S
\ref{sec:Results}, wherein we also examine some special cases
of ring formation in non standard geometries, and connect these to what is observed experimentally. Conclusions are drawn in \S \ref{sec:Conclusion}.

\section{Problem setup and numerical method\label{sec:setup}}

\begin{figure}
\noindent \centering{} \includegraphics[width=0.5\columnwidth]{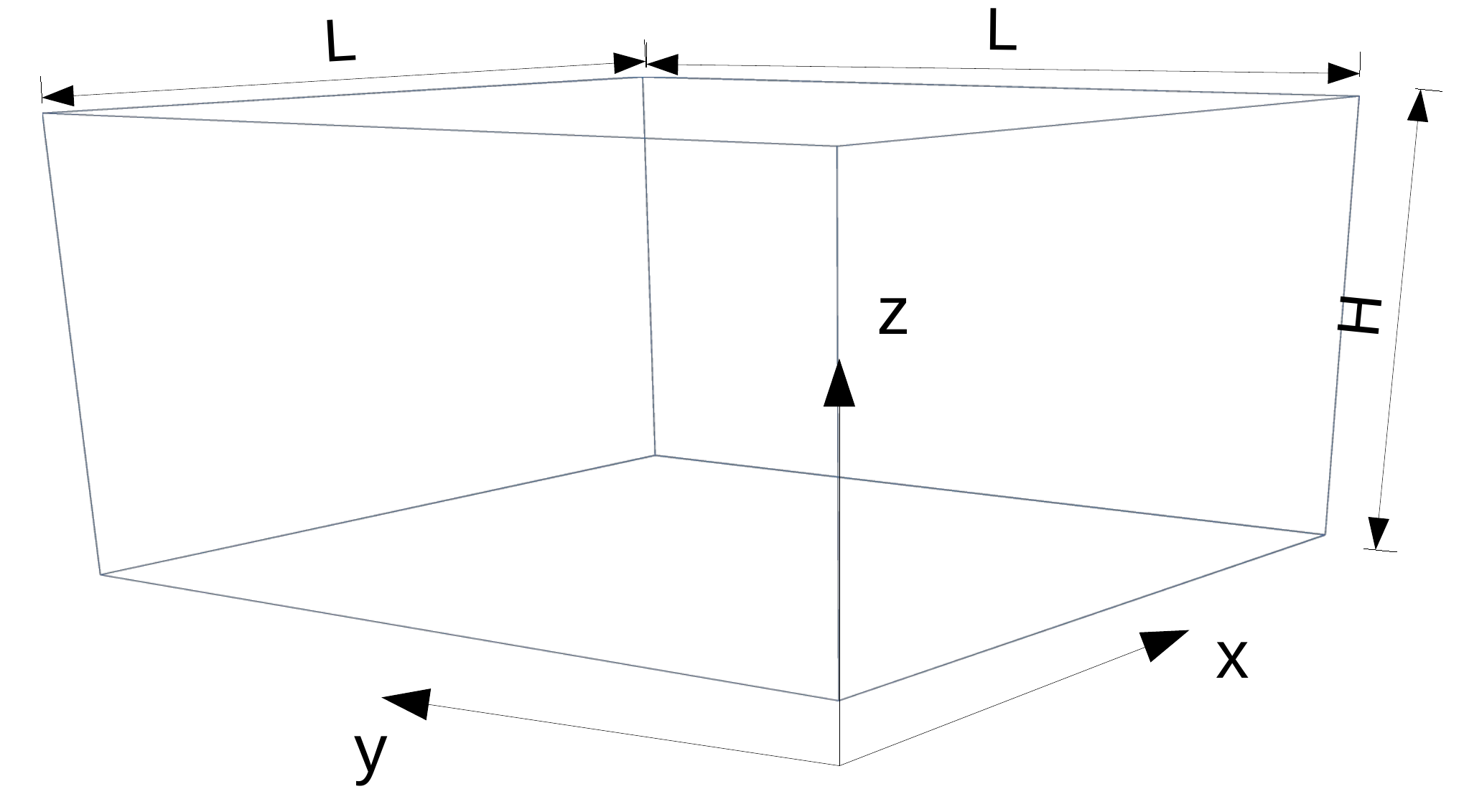}
\caption{The geometry of the problem. The container $[-L/2,L/2]\times[-L/2,L/2]\times[0,H]$
is spun about the vertical z-axis with an angular velocity $\Omega$ starting at $t=0$.}
\label{fig:schematic_box}
\end{figure}

A schematic of the system under study is shown in Figure \ref{fig:schematic_box}.
We consider a container of width $L$ and height $H$ filled with a Boussinesq
fluid of density $\rho$, coefficient of thermal expansion $\alpha$, 
viscosity $\nu$ and thermal diffusivity $\kappa$, at an initial temperature $T_{0}$ that is rotated about the vertical axis
starting at $t=0$. 
In a frame of reference rotating with the container the Coriolis effect is present. The lateral surfaces are thermally insulating and the top and
bottom surfaces have either thermal Dirichlet conditions or one of these horizontal surfaces has a Neumann boundary condition.
Other parameters in the problem include the acceleration
of gravity $-g\boldsymbol{e}_{z}$ (gravity and the axis of rotation
are both along the $z-$axis), and the aspect ratio $A=L/H$ ($=2$
unless otherwise stated). The dimensional equations are
\begin{align}
\frac{D\boldsymbol{u}}{Dt} & =-\frac{\nabla p}{\rho}+\nu\nabla^{2}\boldsymbol{u}-2\Omega\boldsymbol{e}_{z}\times\boldsymbol{u}\nonumber\\
&+g\alpha\left(T-T_{0}\right)\boldsymbol{e}_{z}-\alpha\left(T-T_{0}\right)\Omega^{2}\boldsymbol{r}\label{eq:momentum_dim},\\
\frac{DT}{Dt} & =\kappa\nabla^{2}T\label{eq:temperature_dim} \qquad \textrm{and} \qquad\\
\nabla\cdot\boldsymbol{u} & =0\label{eq:continuity_dim}.
\end{align}
In the rotating frame the initial velocity of the fluid is
\begin{equation}
\boldsymbol{u}|_{t=0}=-\boldsymbol{\Omega}\times\boldsymbol{r},\label{eq:init_cond_dim}
\end{equation}
\tb{and the initial temperature is $\theta=0$ in the entire container. This is different from the fully-developed convective state that is used as the initial condition in the laboratory experiments of \cite{boubnov1986} and \cite{ZPW} prior to spin-up.}

\subsection{Boundary Conditions \label{subsec:BCs}}

The boundary conditions (BCs) for velocity and temperature determine
the nature of the buoyancy forcing and the details of the spin-up process.
In the cuboidal geometry, the six bounding surfaces (the top and bottom
surfaces, and the four lateral boundaries) are impenetrable and thus have
zero normal velocity. Each boundary can have no-slip or
free-slip velocity BCs and Dirichlet ($T=\text{constant}$) or Neumann ($\partial T/\partial n=\text{constant}$) thermal BCs.

We consider here only cases where the lateral surfaces are insulating (i.e. have zero heat flux) and have identical velocity BCs (free-slip or no-slip), 
and the thermal BCs on the top and bottom surfaces are of the same type (either
both Dirichlet or both Neumann). Thus there are eight combinations of BCs.
Of these, the majority of our results are from combinations listed  (in their nondimensionalized form) in Table \ref{tab:BCs} in \S \ref{subsec:BCs} below. Other combinations are mentioned where relevant.

%Henceforth, 
For simplicity we call all boundaries `surfaces', so that, for instance, 
a `free-slip surface' is a boundary where the normal velocity and the
tangential stress are both zero. 

\subsection{Nondimensionalization\label{subsec:Nondimensionalisation}}

We scale time in the problem using the rotation rate, $\Omega^{-1}$,
and the length using the width of the container $L$ (see Figure \ref{fig:schematic_box}).
The choice of $L$ instead of $H$ for the length scale is based on
numerical considerations. These together define the velocity scale
$U=L\Omega$. Assuming a temperature scale $\Delta T$ (to be defined
in the case of constant heat-flux), the governing equations (Eqs. \ref{eq:continuity_dim}-\ref{eq:temperature_dim})
become

\begin{align}
\frac{D\boldsymbol{u}}{Dt} & =-\nabla p+\frac{1}{Re}\nabla^{2}\boldsymbol{u}-2\boldsymbol{e}_{z}\times\boldsymbol{u}+\frac{1}{Fr^{2}}\cdot\theta\boldsymbol{e}_{z},\label{eq:momentum}\\
\frac{D\theta}{Dt} & =\frac{1}{Re\cdot Pr}\nabla^{2}\theta \label{eq:temperature}\qquad \textrm{and} \qquad\\ 
\nabla\cdot\boldsymbol{u} & =0,\label{eq:continuity}\\
\end{align}
where $Pr=\nu/\kappa$ is the Prandtl number, $Re=\Omega L^{2}/\nu$
is the Reynolds number, and $Fr^{-2} = g \alpha \Delta T / \Omega^2 L$ is the Froude number, which 
is a measure of the strength of the buoyancy relative to other
forces. The initial velocity is $\boldsymbol{u}\left(t=0\right)=-\boldsymbol{e}_{z}\times\boldsymbol{r}$
and the initial temperature is $\theta\left(t=0\right)=0$ everywhere
in the container. The BCs are defined in \S \ref{subsec:BCs}. \\

A constant heat flux $\dot{q}$ implies a constant buoyancy flux $\tilde B$,
given in terms of $\dot{q}$ as
\begin{equation}
\tilde{B}=\frac{g\alpha\dot{q}}{\rho C_{p}},\label{eq:buoy_flux_defn}
\end{equation}
where $C_p$ is the heat capacity per unit mass of the fluid at constant pressure.
The flux Rossby number, which is a measure of the buoyancy flux, is
\begin{equation}
Ro_{f}=\sqrt{\frac{\tilde{B}}{\Omega^{3}L^{2}}}, \label{eq:Rof_defn}
\end{equation}
the flux Rayleigh number $Ra_{f}$ is 
%defined in terms of the buoyancy flux as
\begin{equation}
Ra_{f}=\frac{\tilde{B}H^{4}}{\nu\kappa^{2}} \tb{ = \frac{{Ro_f}^2 {Re}^3 {Pr^2}}{A^4}} ,\label{eq:Raf_defn}
\end{equation}
and the Nusselt number is 
\begin{equation}
\tb{Nu=\left\langle{\frac{\theta' H}{\bar{\theta}_{z=0}- \bar{\theta}_{z=H}}}\right\rangle,\label{eq:Nusselt_defn}}
\end{equation}
where $\theta'$ is the constant temperature gradient imposed at $z=H$, \tb{the overbar $\bar{\cdot}$ denotes the spatial average across a given plane}, and  $\langle \cdot \rangle$ denotes the time-average. \tb{For all the results reported here, the time-average was taken over $300<t<600$.}
The standard Rayleigh number follows from the above definitions
and is 
\begin{equation}
Ra=\frac{Ra_{f}}{Nu}.\label{eq:Ra_defn}
\end{equation}
The temperature scale $\Delta T$ is defined as
\begin{equation}
\tb{\Delta T = \frac{\dot q L}{\rho C_p \kappa \theta^\prime}}, \label{eq:deltaT}
\end{equation}
and hence the Froude number can also be written as
\begin{equation}
\tb{Fr^{-2} = \frac{{Ro_f}^2 Re Pr}{\theta^\prime}} . \label{eq:Froude}
\end{equation}
% 
% \tb{When the temperature scale $\Delta T$, as opposed to the heat flux $\dot q$, is fixed by the BCs,
% the above definitions change.

For very large Taylor numbers, $Ta = 4 Re^2 / A^4$, the container of fluid rotates like a solid
body, and for small Taylor numbers, the dynamics resemble non-rotating
Rayleigh-B\'{e}nard convection \citep{boubnov1986}. The boundary between
these is defined by the critical Rayleigh number
\begin{equation}
Ra_{c}\propto Ta^{2/3} \propto Re^{4/3},\label{eq:Rac_defn}
\end{equation}
where the constant of proportionality depends on whether the top- and bottom surfaces obey free-slip or no-slip BCs \citep{boubnov1986}.

\begin{table*}
\noindent \begin{centering}
\begin{tabular}{|c|c|c|c|c|c|c|c|}
\hline 
 {Classification of BCs} & \multicolumn{2}{c|}{Top surface BCs} & \multicolumn{2}{c|}{Bottom surface BCs} & \multicolumn{2}{c|}{Lateral surface BCs} & Rings\tabularnewline
\hline 
 & $\boldsymbol{u}$ & $\theta$ & $\boldsymbol{u}$ & $\theta$ & $\boldsymbol{u}$ & $\theta$ & \tabularnewline
\hline 
Type I (BG, ZPW) & $\partial\boldsymbol{u}/\partial n=0$ & $\partial\theta/\partial n=\theta^\prime$ & $\boldsymbol{u}=0$ & $\partial\theta/\partial n=0$ & $\boldsymbol{u}=0$ & $\partial\theta/\partial n=0$ & Yes\tabularnewline
\hline 
Type II (VE, {*}) & \multirow{1}{*}{$\boldsymbol{u}=0$} & $\theta=-1$ & $\boldsymbol{u}=0$ & $\theta=0$ & $\boldsymbol{u}=0$ & $\partial\theta/\partial n=0$ & Yes\tabularnewline
\hline 
Type III & $\partial\boldsymbol{u}/\partial n=0$ & $\theta=-1$ & $\boldsymbol{u}=0$ & $\theta=0$ & $\boldsymbol{u}=0$ & $\partial\theta/\partial n=0$ & Yes\tabularnewline
\hline 
Type IV & $\boldsymbol{u}=0$ & $\theta=-1$ & $\boldsymbol{u}=0$ & $\theta=0$ & $\boldsymbol{u}=0$ & $\partial\theta/\partial n=0$ & No\tabularnewline
\hline 
Type V & $\partial\boldsymbol{u}/\partial n=0$ & $\theta=0$ & $\boldsymbol{u}=0$ & $\theta=1$ & $\boldsymbol{u}=0$ & $\partial\theta/\partial n=0$ & No\tabularnewline
\hline 
\end{tabular}
\par\end{centering}
\caption{\label{tab:BCs} Combinations of the BCs used for
the results reported. Some other possible combinations are discussed
as special cases in \S \ref{sec:Special_cases}. BCs of Type
I are as used by \cite{boubnov1986} (BG) and \cite{ZPW} (ZPW). Type II is the \cite{vorobieff1998} (VE) setup with a cylindrical
container. Comparing results from Type I and Type III (\S \ref{subsec:Type-I-BCs} and \S \ref{subsec:Type-III-BCs}) elucidates the role of the thermal BCs in the
dynamics. BCs of Types IV and V produce no rings, instead producing
square sheets of convection. 
%$A$ is the aspect ratio, defined in \S \ref{sec:setup}.
}
\end{table*}

\subsection{Numerical Method\label{subsec:megha5}}

The numerical simulations are performed with the finite volume code
\emph{Megha}-5, which uses uniform grids and second order
central differences in space and second order Adams Bashforth timestepping.
The momentum equation is solved using the projection operator \citep{chorin1968}
and the resulting Poisson equation for the pressure is solved using
cosine transforms with the PFFT Library of \cite{PFFT}.
The scalar equation is solved using a local upwind scheme \citep{herrmann2006localupwind}
that avoids Gibbs oscillations while retaining overall second order
accuracy. Alternatively, the second order scheme of \cite{KT2000} can also be used. 
\emph{Megha}-5 is based on an extensively validated earlier version \citep{prasanth2014} and has been used 
in studies of jets and plumes \citep{diwan2014}, \tb{ cumulus \citep{cumulus2020} }and mammatus clouds \citep{ravichandran2020mammatus}.\\

The thickness of the thermal boundary layer adjacent to a surface is defined
as the distance at which the mean temperature of the volume would
be reached starting at the surface temperature with the slope from the
first two gridpoints from the surface, following the convention of \citet{belmonte1994temperature} and \citet{verzicco2008comparison}.
We ensure that the thermal boundary 
layers at the top surface are resolved with at least $6$ gridpoints for 
Reynolds numbers $Re\leq7.5\times10^{3}$ (grid size of $256^2 \times 128$, with a time step of $2.5\times10^{-3}$), and up to $12$ gridpoints (grid size of $512^2 \times 256$, with a time step of $1.25\times10^{-3}$) for $Re\geq10^{4}$, as required in 
turbulent Rayleigh-B\'{e}nard convection \citep[see][and refs. therein]{schell}. The results were found to be grid independent and we report those from the lower resolution grid here.  
We have also verified that the choice of local-upwinding or Kurganov-Tadmor discretization
does not affect the results (the former is used unless otherwise mentioned).

Simulations in the cylindrical geometry and other geometries mentioned
in \S \ref{sec:Special_cases} are performed using the volume
penalization method \citep{kevlahan2001computation,schneider2005numerical},
with insulating BCs for the simulations in the cylindrical
geometry applied following \cite{Kadoch2012}. Our results are
verified to be independent of the penalization parameters used. 

\section{Results and Discussion \label{sec:Results}}

\begin{figure}
\noindent \begin{centering}
\includegraphics[width=1\columnwidth]{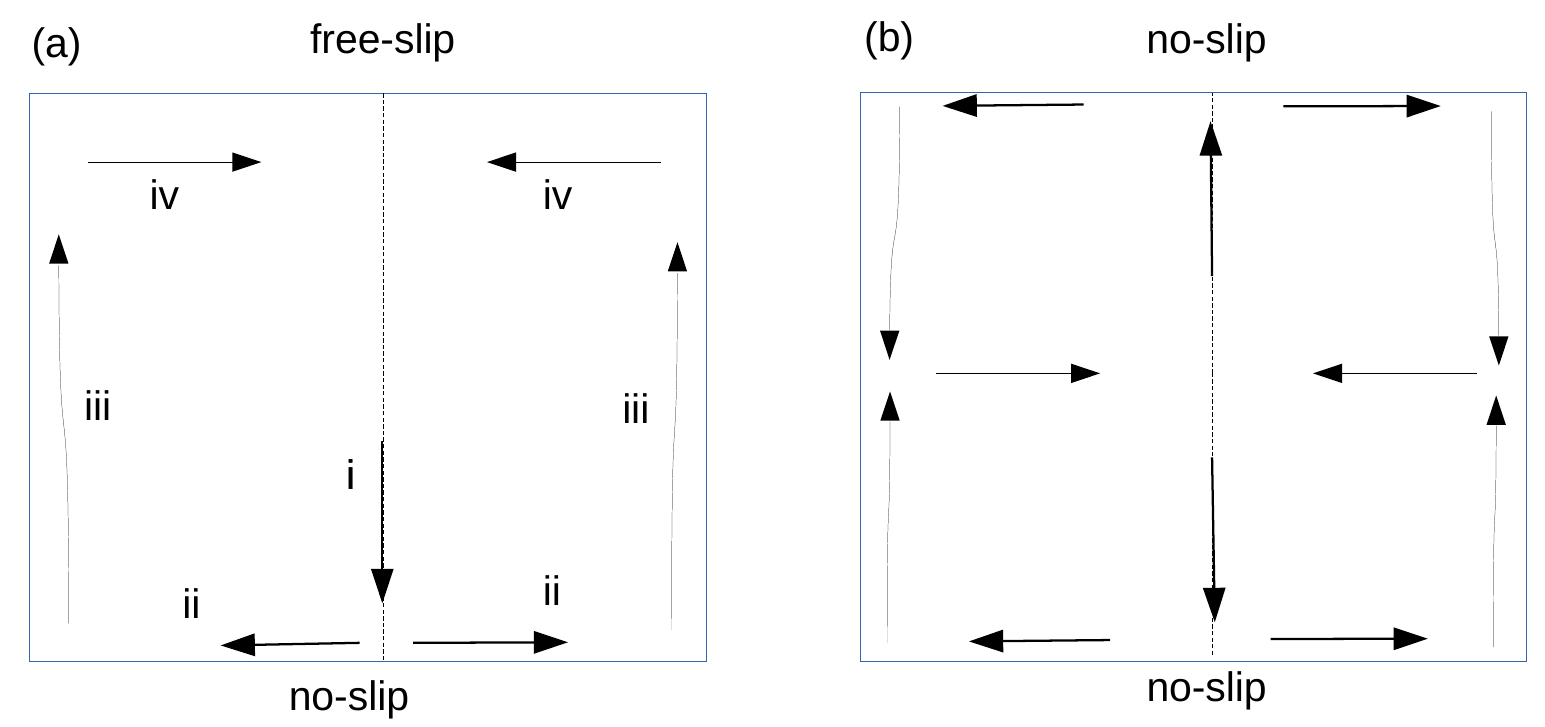}
\par\end{centering}
\caption{\label{fig:schematic_freeslip_vs_noslip} Schematic vertical sections
through the plane of symmetry of the flow during convective spin-up
with (a) free-slip top boundary, and all other surfaces no-slip; and
(b) all surfaces no-slip. The arrows show flow directions (magnitudes
not to scale). {The central dashed line is the axis of rotation (and the direction of gravity). 
The four labels correspond to (i) flow towards the no-slip surface;
(ii) centrifugally outwards flow at a no-slip surface; (iii) flow rising vertically along
the lateral boundaries; and (iv) return flow towards the axis of rotation. This pattern and
its mirror-image are seen in (b).}}
\end{figure}

We begin by summarising the spin-up process in the absence of buoyancy
forcing, following GH. Consider the case where
the top surface is free-slip and the bottom surface and the four lateral
boundaries are no-slip surfaces, as shown schematically in Figure \ref{fig:schematic_freeslip_vs_noslip}(a).
The flow at the bottom surface is that due to a plate
impulsively rotated about an axis perpendicular to its plane \citep[see e.g., Chapter 5.2.4, p. 119 of][]{schlichting2016boundary}. Fluid
is centrifuged outwards from the axis of rotation along the surface.
Continuity drives fluid downward towards the bottom surface. As the centrifuged
fluid reaches the periphery of the container, it ascends up the lateral surfaces,
driven by a vorticity gradient that exists as a result of the boundary
layers on the lateral surfaces. Once this fluid reaches the upper free surface,
it is driven towards the axis, eventually becoming part of the downward flow. 
In this manner, fluid is driven from larger to smaller radii. Conservation of angular momentum
(excepting for small viscous losses) insures that the fluid near
the axis is replaced with fluid that is rotating more rapidly. GH show that 
this process takes a time $\mathcal{O} \left(\Omega^{-1} Re^{1/2} \right) = \mathcal{O} \sqrt{ \left( L^2 / \nu \Omega \right) } $.

\subsection{Type I BCs \label{subsec:Type-I-BCs}}

\begin{figure}
\begin{singlespace}
\noindent \begin{centering}
\includegraphics[width=0.5\columnwidth]{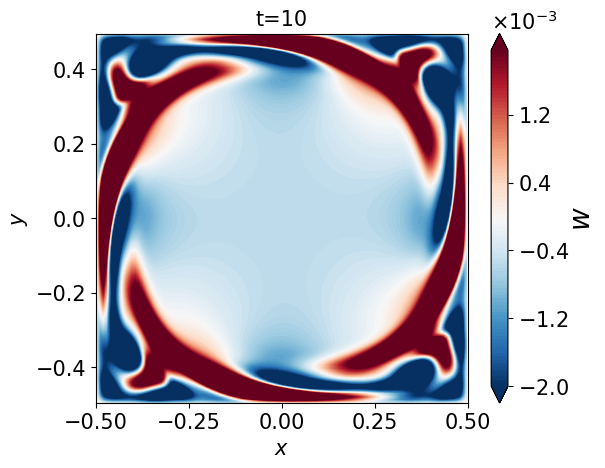}\includegraphics[width=0.5\columnwidth]{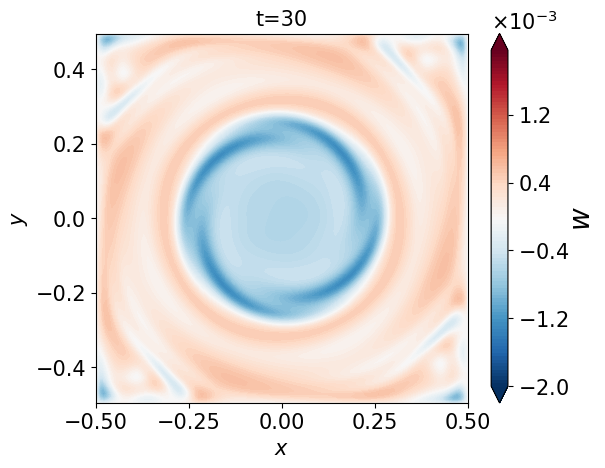}
\par\end{centering}
\end{singlespace}
\noindent \begin{centering}
\includegraphics[width=0.5\columnwidth]{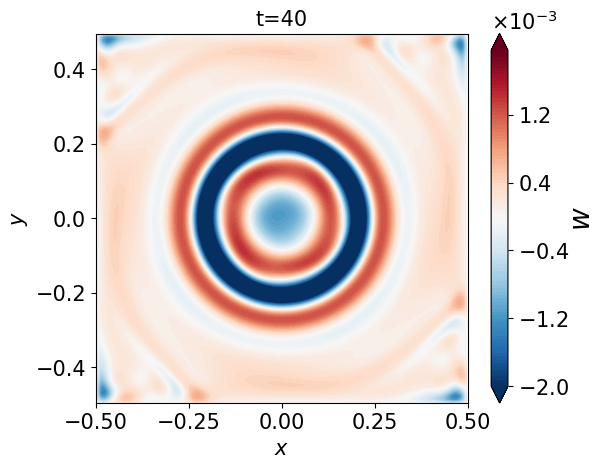}\includegraphics[width=0.5\columnwidth]{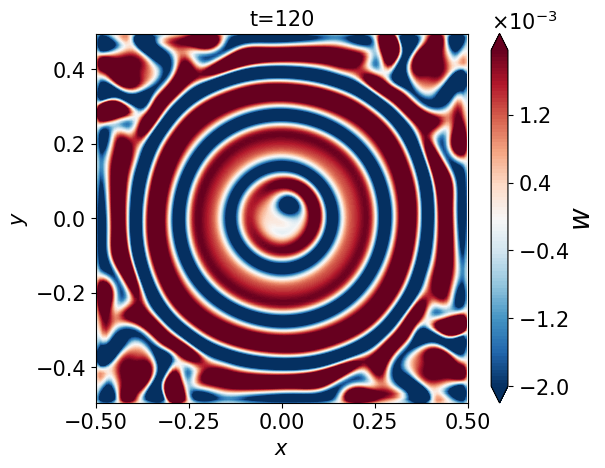}
\par\end{centering}
\begin{singlespace}
\noindent \begin{centering}
\includegraphics[width=0.5\columnwidth]{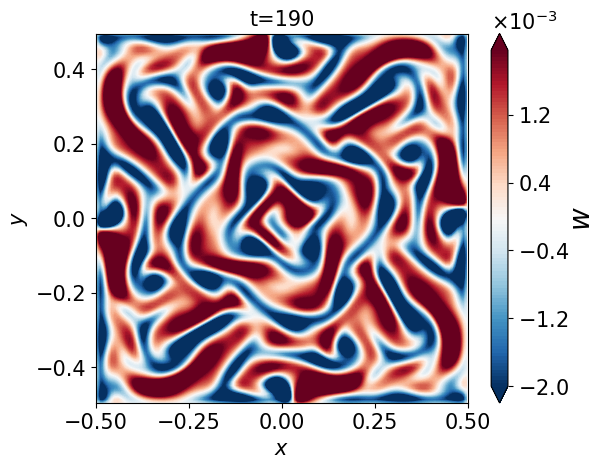}\includegraphics[width=0.5\columnwidth]{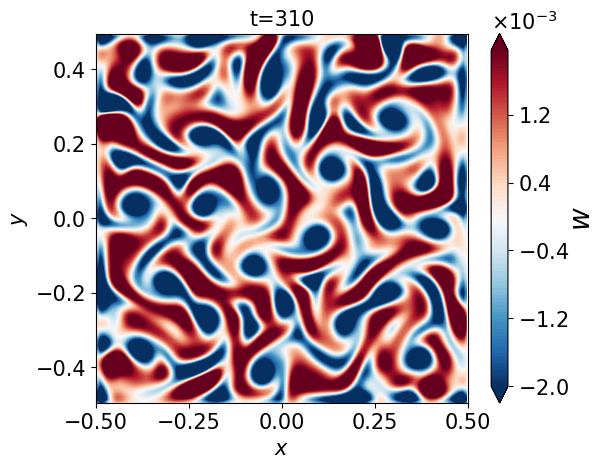}
\par\end{centering}
\end{singlespace}
\caption{\label{fig:type_I_rings} Ring formation in a representative case
for Type I BCs. Shown are horizontal cross-sections
of the vertical velocity field at a plane $z\approx0.47$ which is
near the cooled upper surface. The parameters of the problem are $Re=7500$,
$Ro_{f}=0.00442$, $Pr=5$. (See \tb{Figure \ref{fig:Hovmoller} for a Hovm\"oller plot showing the time-evolution and} Figure \ref{fig:compare_type_III_typeI} for a sequence of vertical cross-sections.) \tb{The evolution for these fields is available as a movie in the supplementary material.}}
\end{figure}

\begin{figure}
\noindent \begin{centering}
\includegraphics[width=0.6\columnwidth]{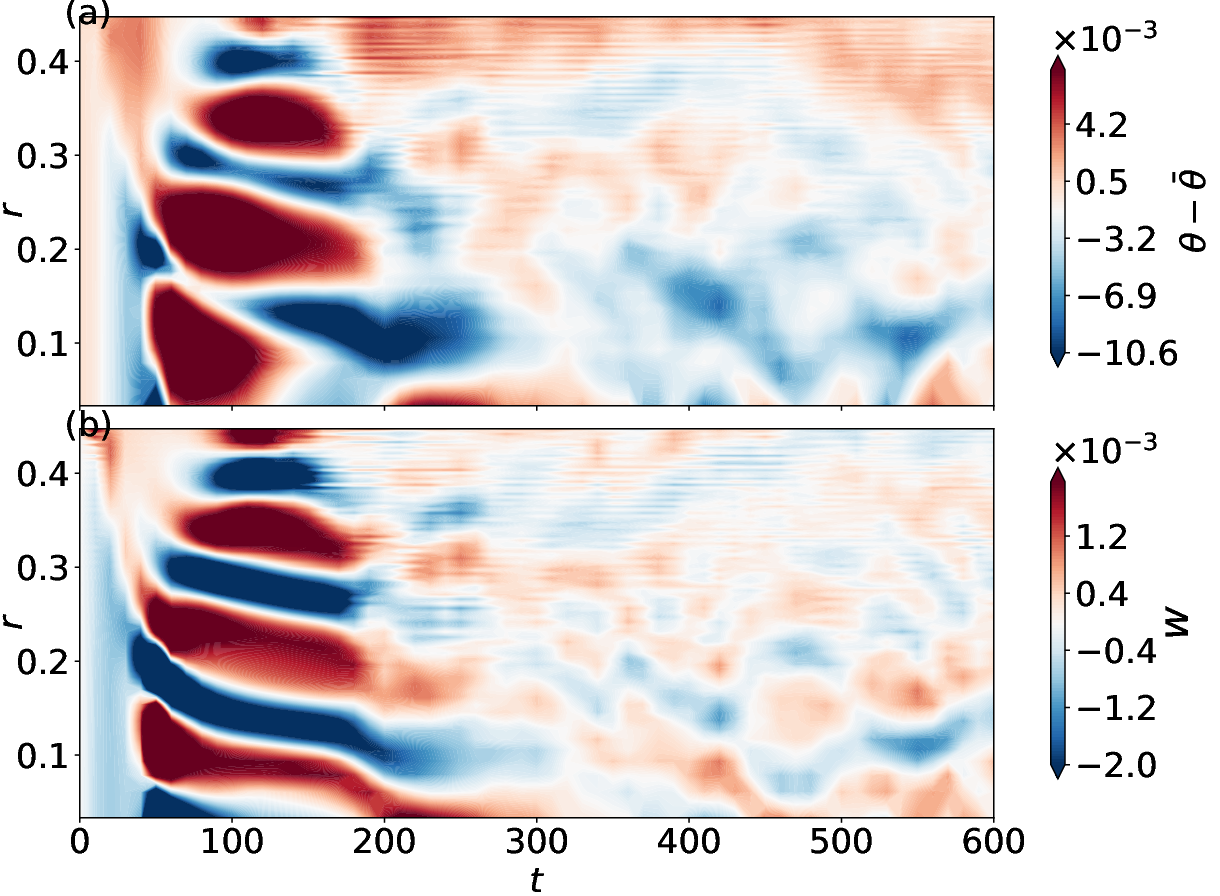}
\par\end{centering}
\caption{\label{fig:Hovmoller} \tb{Hovm\"oller plots for (a) the temperature difference $\theta - \bar{\theta}$, and (b) the azimuthally averaged vertical velocity $\textrm{w}$, where  $\bar{\theta}$ is the average temperature in the plane $z=0.46$ where the plots are made, showing the evolution of the rings. The rings can be seen to form around $t=40$, move radially inwards, and break down around $t=150$ coinciding with the completion of spin-up. $Re=7500, Ro_f=0.00442, Pr=5$, as in Fig. \ref{fig:type_I_rings}.}}
\end{figure}

We first discuss results from simulations with Type I BCs (see Table \ref{tab:BCs}); the sides and the bottom are all no-slip, thermally insulating surfaces and the upper free-slip surface is driven with a constant heat flux.  
The dynamical time scale for the
circulation shown schematically in Figure \ref{fig:schematic_freeslip_vs_noslip}(a) is fast
relative to the build up of negatively buoyant fluid at the upper surface. As
cold plumes emerge, they are sequentially forced towards the axis of rotation 
as buoyancy and rotational forces balance, the oldest and more central of which
are deeper. A given plume evolves into an axisymmetric ring as this quasi-steady
balance is attained, thereby leading to a sequence of upwelling and downwelling
ring pairs. Up to three pairs are seen for such $Re \leq 10^4$. The rings eventually reach the bottom of the box, where they interact with the boundary layer and are influenced by the shape of the container if a sufficiently long time passes. As the system approaches solid-body rotation, the system must become unstable and break up into cyclonic vortices, in which fluid sinks surrounded by regions of slower upwelling flow. While this generic process remains similar across a wide parameter range, the ring and vortex numbers are a function of the Reynolds, flux Rossby and Prandtl numbers. A sequence of images showing this evolution is presented in Figure \ref{fig:type_I_rings}, \tb{ and Hovm\"oller plots showing the evolution of the azimuthally averaged vertical velocity and temperature are shown in Fig. \ref{fig:Hovmoller}.}

\begin{figure}
\noindent \begin{centering}
\includegraphics[width=1\columnwidth]{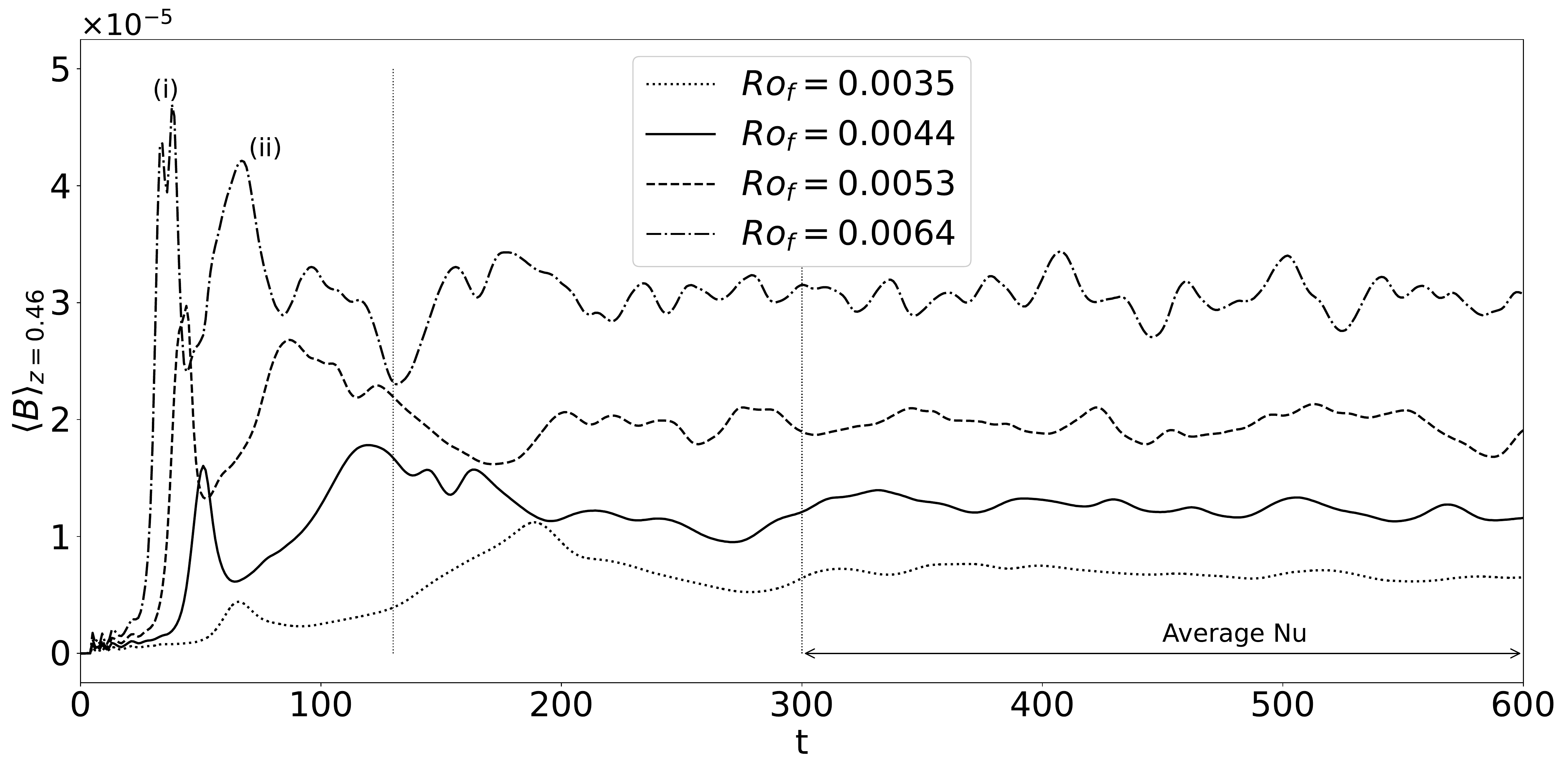}
\par\end{centering}
\caption{\label{fig:buoy_flux_typeI} The buoyancy flux at $z=0.455$ as a
function of time, for $Re=7500$ and $Pr=5$. The peaks in buoyancy flux, 
labelled (i) and (ii) in the figure, correspond to the time when the first 
ring forms and \tb{(less precisely to)} the time at which the ringed state is maximal respectively.
\tb{The two dotted vertical lines are at $t=130$, approximately when spin-up is complete and the rings
start to break down; and $t=300$, when we begin calculations of time-averages. 
Increasing the Rossby number increases the buoyancy flux imposed at the
upper surface. The stages of ring formation (maximal state and breakdown) occur
earlier when the flux Rossby number is larger. } }
\end{figure}

To show the heat transport by the rings, we plot the cross sectional area-averaged dimensionless buoyancy flux, defined as
\begin{equation}
\langle B \rangle \left(z,t\right) =\frac{1}{Fr^{2}}\int\limits_{-1/2}^{1/2}\int\limits_{-1/2}^{1/2} dxdy\left( \textrm{w} \theta\right),\label{eq:buoy_flux_wtheta}
\end{equation}
at a horizontal section at $z=0.455$. 
% In the above,
% \begin{equation}
% Fr^{-2} = \frac{g\alpha \Delta T}{\Omega^2 L} \label{eq:defn_Fr2} 
% \end{equation}
% is the Froude number, a ratio of buoyancy to rotational forces, and depends on the nature of the thermal forcing.
% For Type I BCs, with the temperature scale $\Delta T$ as defined above, the Froude number becomes
% $Fr^{-2}_{I} = {Ro_f}^2 Re Pr$.\\
The first two peaks of buoyancy flux seen in Figure \ref{fig:buoy_flux_typeI} correspond
to the formation of the first ring and the maximally ringed state respectively. \\

In the limit of very small $Ro_{f}$ (Eq. \ref{eq:Rof_defn}), convection is strongly suppressed. For large $Ro_{f}$, the heat flux dominates the effects
of rotation. The dynamics of ring formation are most prominent
at \sout{some} intermediate value\tb{s} of \tb{$Ro_{f}$}, as found by \cite{boubnov1986} and \cite{ZPW}. As $Ro_{f}$ increases, the time for the first ring to form decreases and its radius increases.\\

% \begin{figure}
% \noindent \begin{centering}
% \includegraphics[width=0.5\columnwidth]{figs2/AR2_const_flux/standard/firstring_vs_Rof_Pr5}
% \par\end{centering}
% \caption{\label{fig:ring_locs} The radius of the first ring as a function
% of $Ro_{f}$ for different $Re$. $Pr=5$. The radius is plotted at the maximally ringed state.
% For the larger of the Reynolds and flux Rossby numbers, the inner rings have disintegrated,
% leading to the up-and-down behaviour.}
% \end{figure}

% \sout{For low $Pr$, coherent rings dissipate rapidly and for high $Pr$\tb{, as the rings become thinner}, the increased inter-ring shear drives instability the widths of plumes that emerge from the top boundary layer are inversely proportional to $Pr$. }.

\begin{figure}
\noindent \begin{centering}
\includegraphics[width=1\columnwidth]{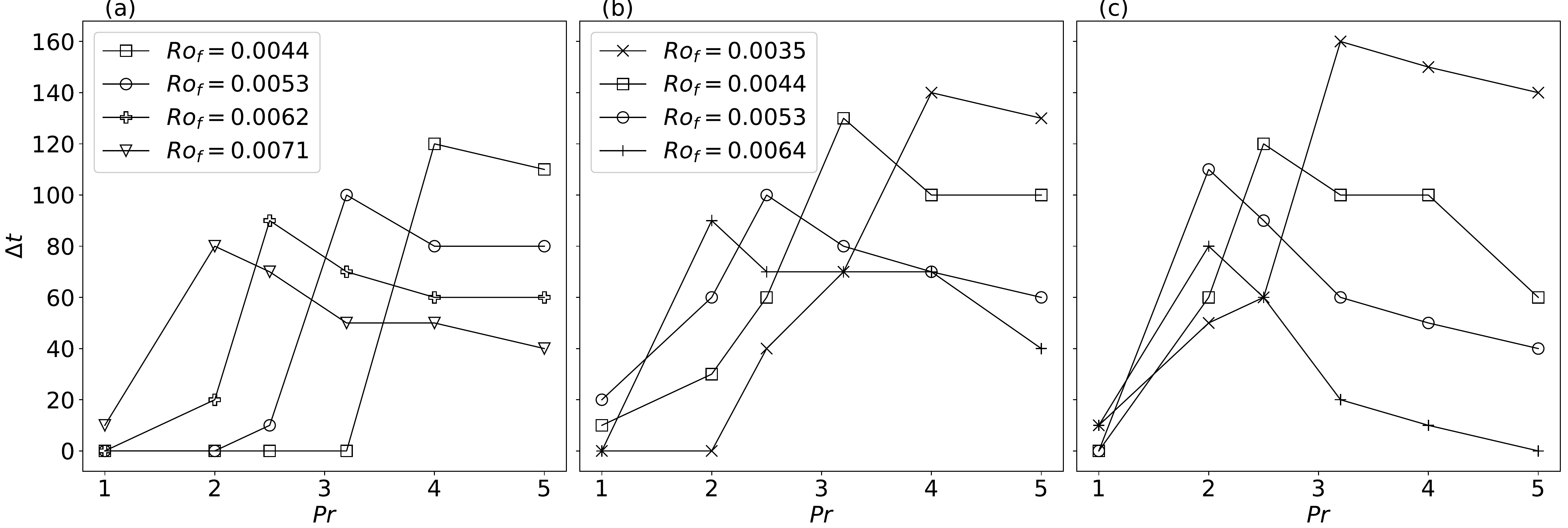}
\par\end{centering}
\caption{\label{fig:longevity_vs_Pr} The longevity of the ringed state for
three different Reynolds numbers and various flux Rossby numbers,
showing the variation of $Pr^{*}\left(Re,Ro_{f}\right)$ for 
(a): $Re=5000$, (b): $Re=7500$, (c): $Re=10000$. The legends
show the values of $Ro_{f}$ for which the lifetimes are plotted. The legends
for $Re=7500$ and $Re=10000$ are shared. \tb{For $Re=5000$ and $Ro_f=0.0044$, rings
do not form for $Pr\leq3.2$.}}
\end{figure}

\begin{figure}
\noindent \begin{centering}
\includegraphics[width=0.5\columnwidth]{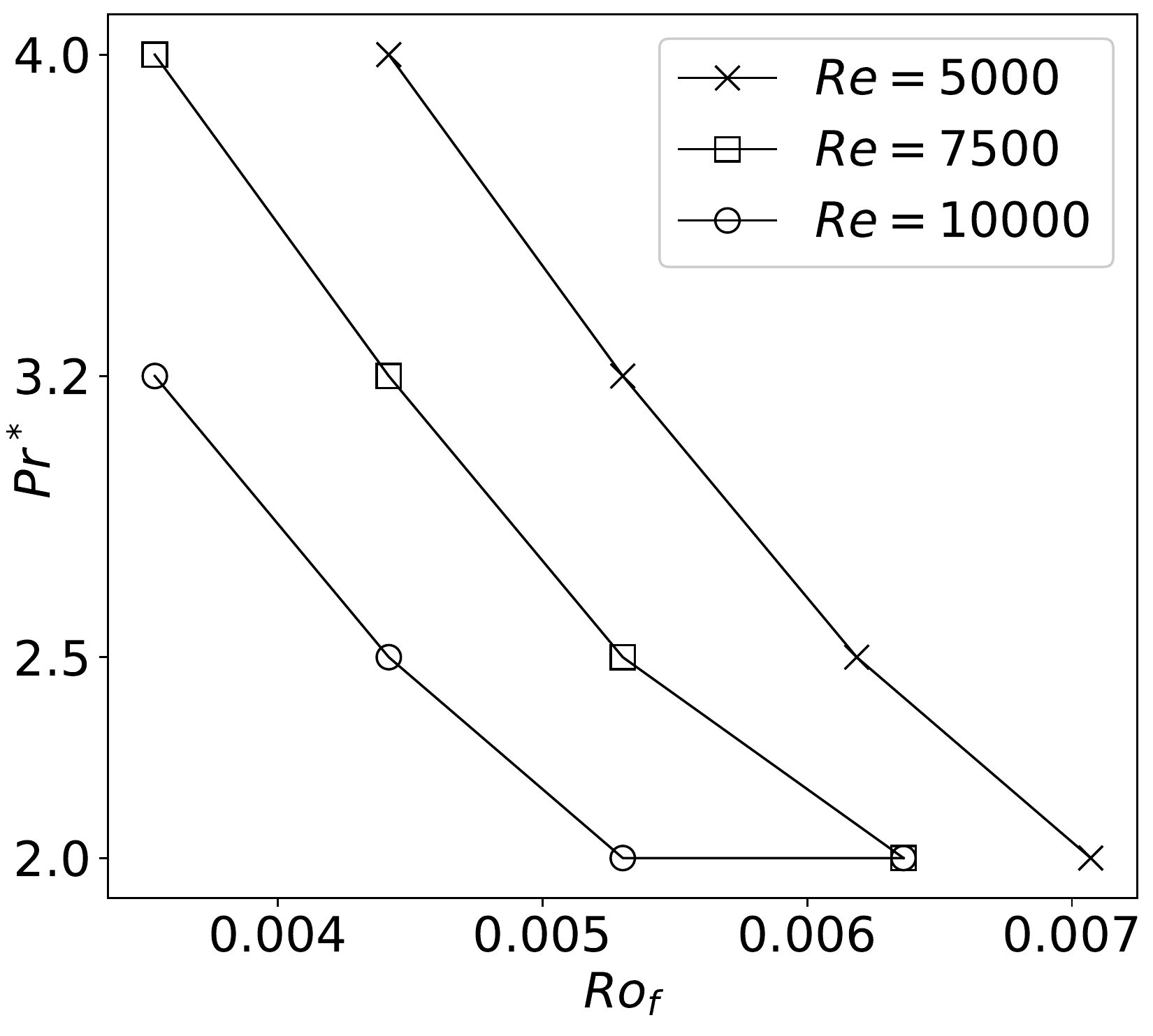}
\par\end{centering}
\caption{\label{fig:Prc} The Prandtl number $Pr^{*}\left(Re,Ro_{f}\right)$ at which the rings are the longest-lived.  \tb{The $Pr^*$ for $Ro_f=0.0064$ appears to be the same for $Re=7500$ and $Re=10000$ because of limited resolution in $Pr$.} See also Fig. \ref{fig:longevity_vs_Pr}.}
\end{figure}

The Prandtl number strongly influences the dynamics--particularly the stability of the ringed state. For a given $Re$, \tb{as thermal dissipation is decreases and $Pr$ increases, the rings become thinner. Thus, while the thermal effect increases the ring longevity with $Pr$, the associated thinning of the rings enhances the across-ring shear, driving the shear instability (note that the flow turns in opposite directions on either side of a ring) and thereby reducing the ring longevity}. Thus, the stability of the rings peaks at an intermediate Prandtl number $Pr^{*}$. The parameter $\Phi$ measures deviations from axisymmetry of a flow-variable $\phi$ as 
\begin{align}
\Phi & =\int\limits_{0}^{r_\mathrm{max}} dr\left[\phi\left(x,y,z_0,t\right)-\phi(r,z_0,t)\right]^{2} \label{eq:Phi},
\end{align}
where $r_\mathrm{max}=0.45$ and $\phi\left(r,t\right)$ is the average value at radius $r$ at time $t$, and $z_0=0.47$. When $\Phi \leq \Phi_b \left(t = t_b \right)$, we can define the longevity of the ringed state as $t_b$. Figure \ref{fig:longevity_vs_Pr} shows the variation of the lifetime of the ringed state with the system parameters. (Clearly $\Phi$ is also zero if $\phi=0$ everywhere. Thus, a threshold for $\Phi$ is used.)  It can be seen that $Pr^{*}$ is a decreasing function of $Re$ and $Ro_{f}$, as shown in Figure \ref{fig:Prc}. 

The ringed state breaks down into columnar vortices at a time $t_{\text{breakdown}}\left(Re,Ro_{f},Pr\right)$
that follows the second buoyancy flux peak as seen in
Figures \ref{fig:buoy_flux_typeI} and \ref{fig:buoy_flux_typeIII}.

\begin{figure}
\includegraphics[width=0.98\columnwidth]{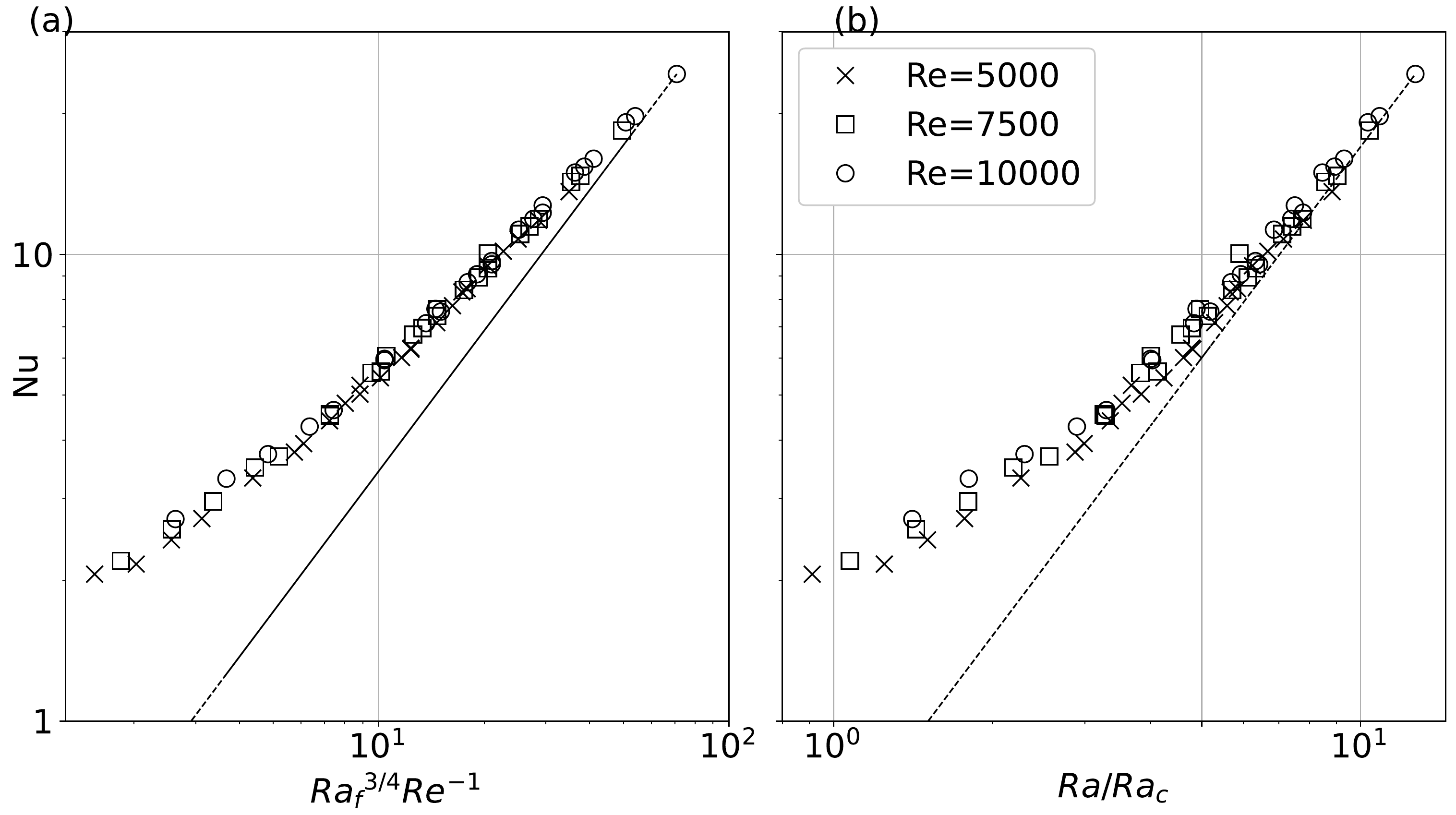}
\caption{\label{fig:Nu_vs_Ra} (a) \tb{The Nusselt number as a function of the appropriately scaled flux Rayleigh number (Eq. \ref{eq:Nu_vs_Ra}). The line is $Nu \sim {Ra_f}^{3/4} {Re}^{-1}$. (b) The Nusselt number as a function of $ Ra/Ra_c$, where $Ra_c$ is the critical Rayleigh number of Eq. \ref{eq:Nu_vs_Ra_by_Rac}, which is the dashed line. 
%\tr{In both figures, the sizes of the markers denote the Prandtl number.} 
We cannot fit power laws with this range of data, but we note the collapse of the simulation data itself in these scalings with that of \cite{Julien2012} in (b) converging to the large $Ra/Ra_c$ behavior \tb{for $Ra/Ra_c \gtrsim 4$.  See also Fig.~ \ref{fig:Nu_compensated_vs_RaE3by2}(b).}}} \end{figure}

\begin{figure}
\includegraphics[width=0.98\columnwidth]{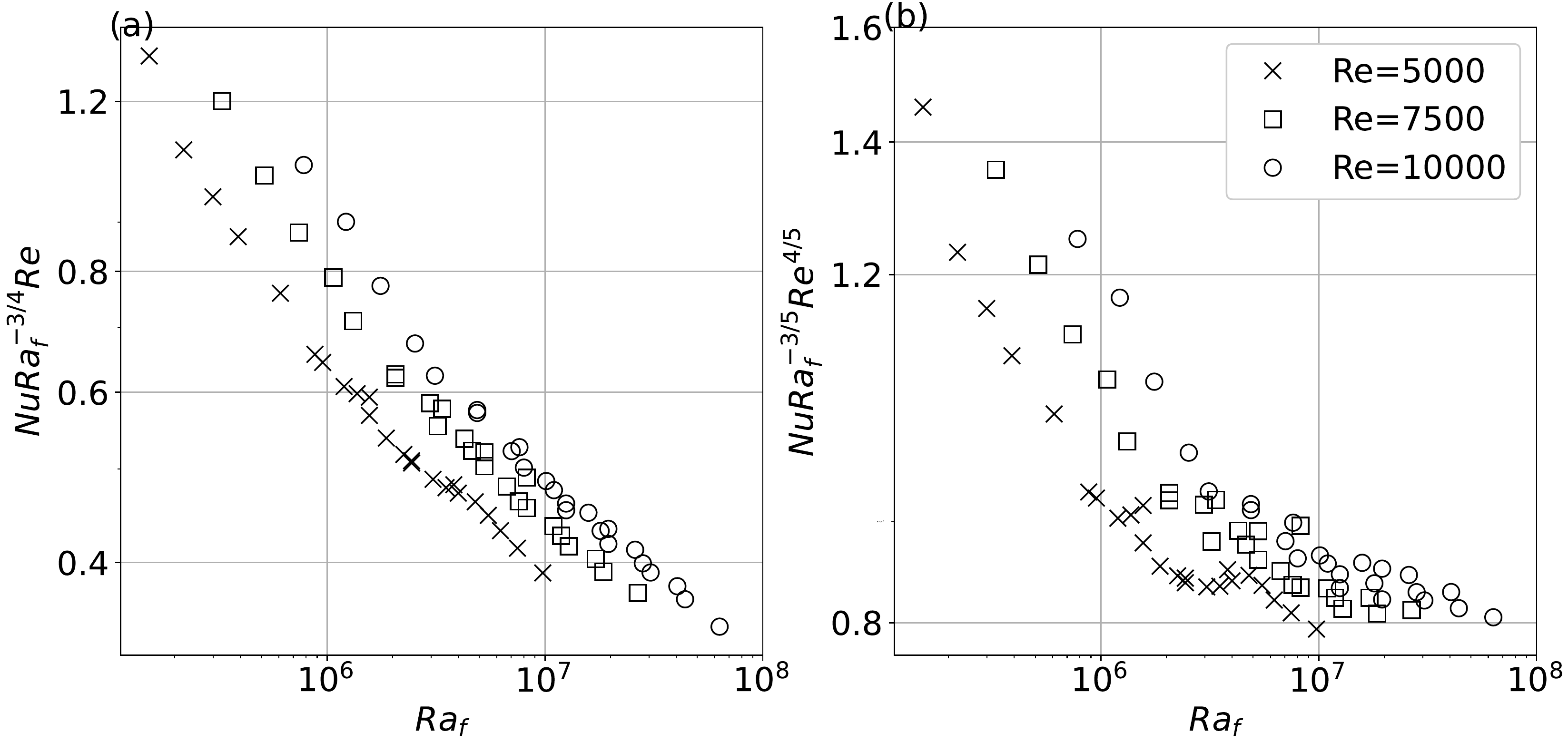}
\caption{ \tb{\label{fig:Nu_compensated} (a) The Nusselt number compensated following Eq. \ref{eq:Nu_vs_Ra} as a function of the flux Rayleigh number. (b) The Nusselt number compensated following Eq. \ref{eq:Nu_vs_Ra_by_Rac} as a function of the flux Rayleigh number. }}
%\tr{In both figures, the sizes of the markers denote the Prandtl number.}} 
\end{figure}

\begin{figure}
\includegraphics[width=1\columnwidth]{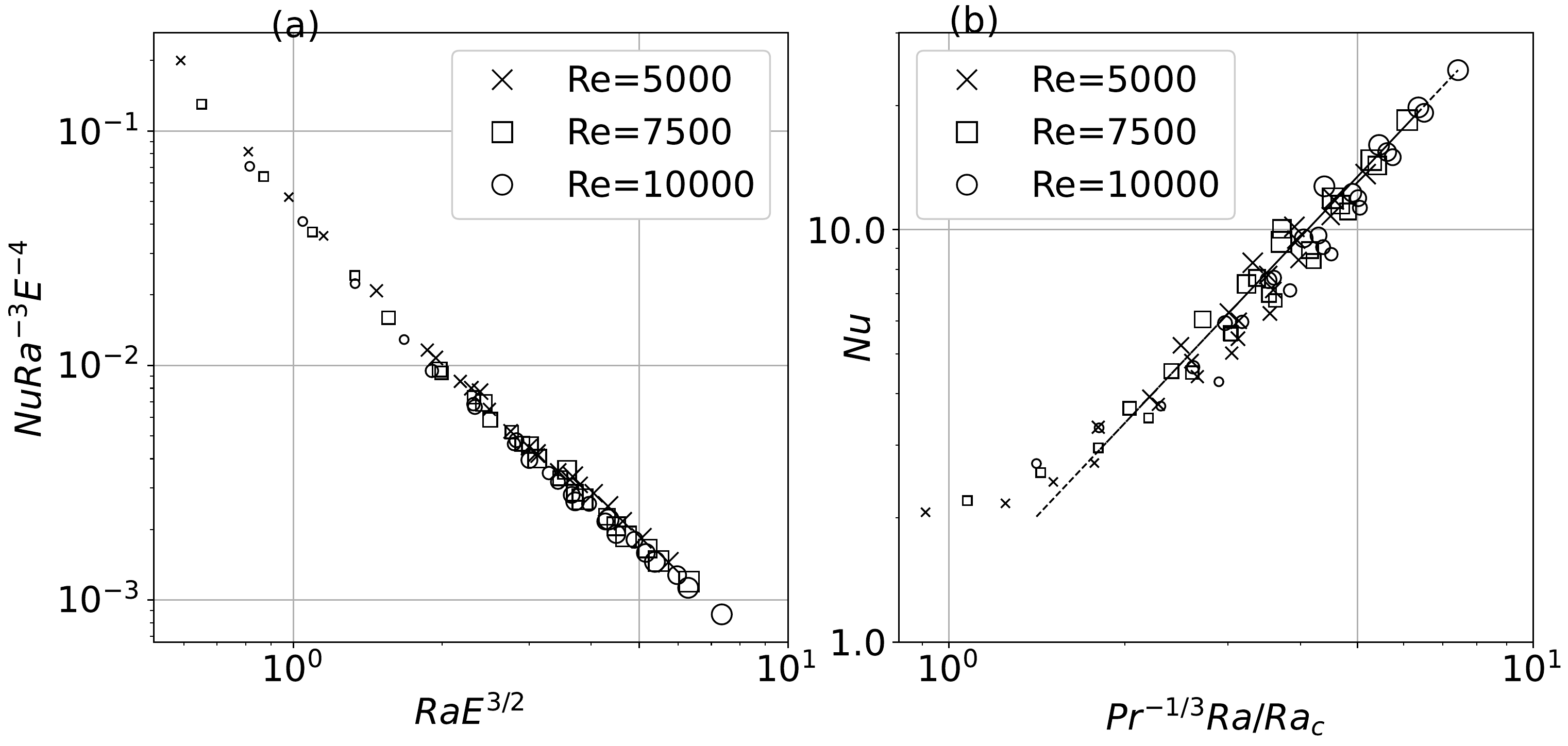}
\caption{ \tb{\label{fig:Nu_compensated_vs_RaE3by2} (a) The compensated Nusselt number, as in Fig. 6(b) of \cite{King2012}.  The absence of an Ekman layer, and thus the absence of a crossover of the thermal and Ekman boundary layers, is responsible for the lack of crossover here, as opposed to that found in Fig. 6(b) of \cite{King2012} and the deviation from scaling in Fig. \ref{fig:Nu_vs_Ra}(a).  (b) With the Prandtl number dependent prefactor to Eq. \ref{eq:Nu_vs_Ra_by_Rac}, as in Eq. 2 of \cite{Julien2012}. In both figures, the marker sizes from small to large correspond to $Pr=(1,2,2.5,3.2,4,5)$ respectively, and increase $\propto \sqrt{Pr}$. }}
\end{figure}

\textcolor{black}{The steady state Nusselt number is calculated from Eq. (\ref{eq:Nusselt_defn}) as a function of the other parameters in the system.  In Figure \ref{fig:Nu_vs_Ra} we show that the simulations collapse to a single curve for different $Re$ when plotted with scalings that emerge from two different treatments of rotating convection.  First, because $Nu \propto {Ra^3}{Ta^{-2}} \propto {Ra^3}{Re^{-4}} = {{Ra_f}^3}{{Nu}^{-3} Re^{-4}}$ 
in geostrophic convection  \citep{boubnov1990, King2012}, one finds that
\begin{equation}
Nu \propto {Ra_f}^{3/4} {Re}^{-1}\label{eq:Nu_vs_Ra}, 
\end{equation}
which is shown in Figure \ref{fig:Nu_vs_Ra}(a) along with the simulation results.  Second, an alternate scaling for the Nusselt number proposed by \cite{Julien2012} is
\begin{equation}
Nu \propto  (Ra/Ra_c)^{3/2} \label{eq:Nu_vs_Ra_by_Rac},
\end{equation}
with $Ra_c = 2.39 {Ta}^{2/3}$ \citep{boubnov1990} which, as shown in Figure \ref{fig:Nu_vs_Ra}(b), captures a larger range of the simulation results. 
Given the small range of the abscissa, we cannot justify fitting power laws, but another means of observing how the simulation results compare to these scalings is using compensated plots as follows.  The appropriate compensated plot for  Eq. \ref{eq:Nu_vs_Ra} is $Nu Re {Ra_f}^{-3/4}$ vs. ${Ra_f}$ and for Eq. \ref{eq:Nu_vs_Ra_by_Rac} is $Nu  {Re}^{4/5}{Ra_f}^{-3/5}$ vs. ${Ra_f}$ as shown in Figures \ref{fig:Nu_compensated}(a) and \ref{fig:Nu_compensated}(b) respectively.  Without manipulation of the prefactor, the latter shows slopes approaching scaling over a wide range of $Re$ for \tb{$Ra_f \gtrsim 5\times10^6$  }. Clearly, this motivates simulations and experiments for an expanded range of $Ra_f$. \tb{A consequence of the arguments used in deriving the scaling in Eq. \ref{eq:Nu_vs_Ra}, which originate in Rossby's interpretation of his experimental data \citep{Rossby1969}, 
is that the Nusselt number curve changes slope when the thermal and Ekman boundary layers cross over and $Ra E^{3/2} = \mathcal{O}(1)$. However, since the upper boundary is one of free-slip and has no Ekman layer, this argument of \cite{King2012}, first articulated by \cite{Rossby1969}, is not operative in this situation, as shown in Fig. \ref{fig:Nu_compensated_vs_RaE3by2}(a). The Prandtl number correction to Eq. \ref{eq:Nu_vs_Ra_by_Rac} given by \cite{Julien2012} is shown in \ref{fig:Nu_compensated_vs_RaE3by2}(b)}.}
\tb{We note that our parameters are comparable to those at which the Nusselt numbers would be expected to increase with rotation \citep{Rossby1969}, were it not for the fact that the upper surface is one of free-slip (Fig.~\ref{fig:schematic_freeslip_vs_noslip})}

\subsection{Type II and Type IV BCs \label{subsec:Type-II-BCs}}

The results of \S \ref{subsec:Type-I-BCs} are qualitatively similar to experiments of \cite{boubnov1986} and \cite{ZPW}
because Type I BCs are similar to the experimental BCs, which have free upper surfaces that are cooled by the evaporation of water. 
\cite{boubnov1986} comment that they observe no rings if the cooled top surface is one of no-slip. Since \cite{boubnov1986} report
experiments in both square-cross-sectioned and cylindrical containers,
it was presumed that they meant this for both geometries. 
\cite{vorobieff1998} perform experiments in cylindrical containers and their rings eventually break up into
vortices as in the cylindrical geometry experiments of \cite{boubnov1986}, but they form much further away
from the axis of rotation. The first ring forms close to the outer lateral surface.  

\begin{figure}
\begin{singlespace}
\noindent \begin{centering}
\includegraphics[width=0.5\columnwidth]{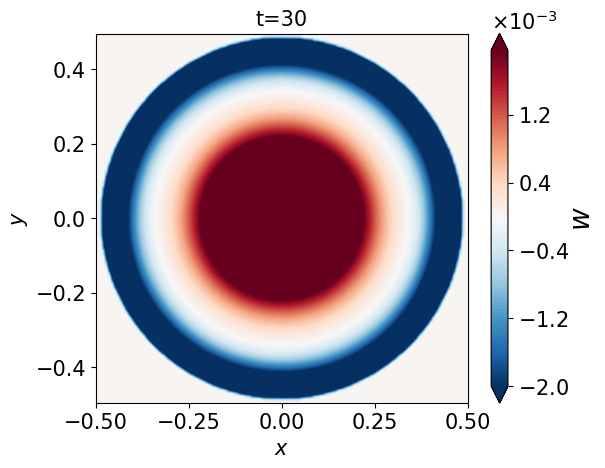}\includegraphics[width=0.5\columnwidth]{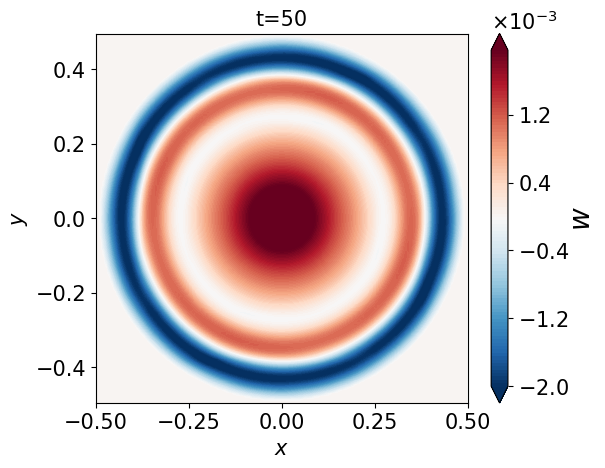}
\par\end{centering}
\noindent \begin{centering}
\includegraphics[width=0.5\columnwidth]{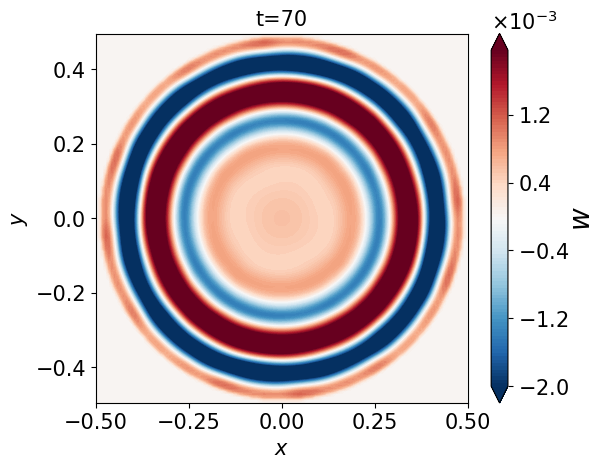}\includegraphics[width=0.5\columnwidth]{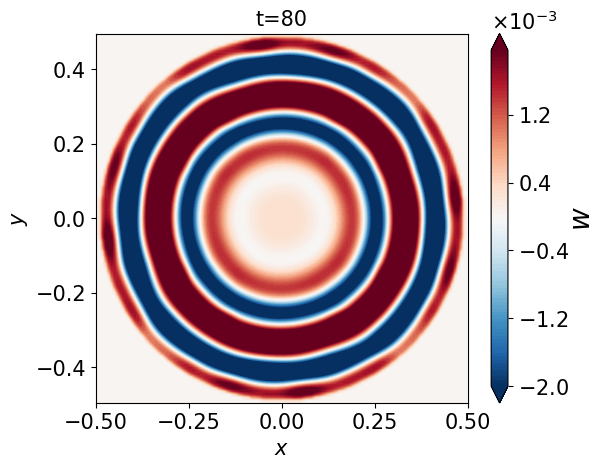}
\par\end{centering}
\end{singlespace}
\caption{\label{fig:type_II_rings} Ring formation for the Vorobieff-Ecke (Type II) BCs. 
Shown are horizontal cross-sections of the vertical velocity field $\textrm{w}$ at $z=0.47$ at $t=30,50,70,80$. 
The parameters of the problem are $Re=5000$, $Ro=0.04$, $Pr=5$. The cylindrical geometry
is embedded in the Cartesian grid using the volume-penalization method (see \S \ref{sec:setup}) }
\end{figure}

We implement the cylindrical geometry using volume penalization,
as discussed in \S \ref{sec:setup}. A sequence of images showing
the evolution for a particular case is shown in Figure \ref{fig:type_II_rings},
which may be compared with that in Figure \ref{fig:type_I_rings}.

\begin{figure}
\noindent \begin{centering}
\includegraphics[width=0.5\columnwidth]{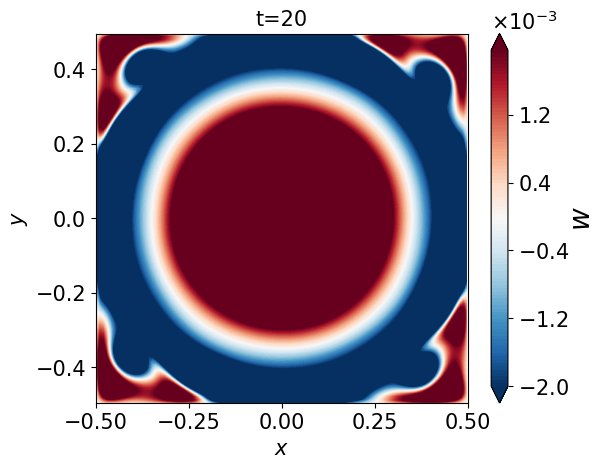}\includegraphics[width=0.5\columnwidth]{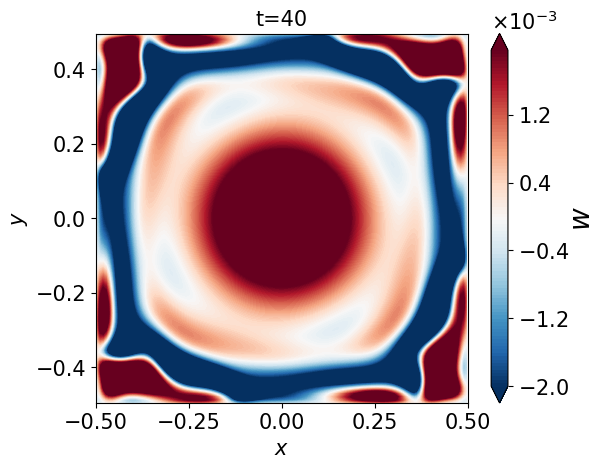}
\par\end{centering}
\noindent \begin{centering}
\includegraphics[width=0.5\columnwidth]{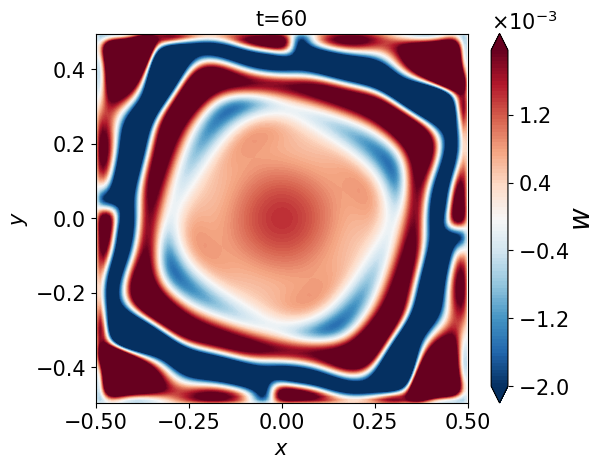}\includegraphics[width=0.5\columnwidth]{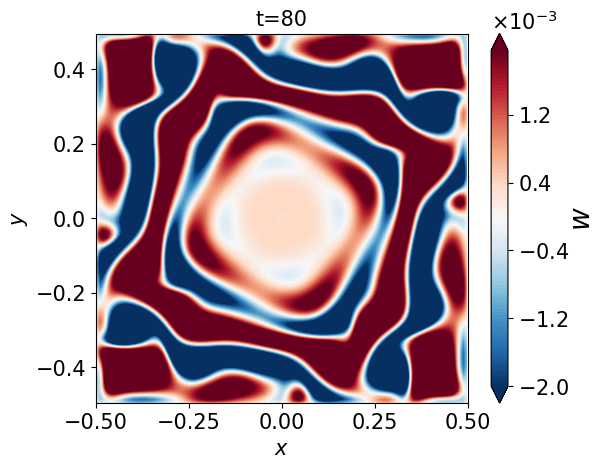}
\par\end{centering}
\caption{\label{fig:typeIV_sheets} Sheet-like convection for Type IV BCs (no-slip
top and bottom surfaces). $Re=5000,Ro=0.04,Pr=5$. \tb{Cross-sections of the vertical velocity} are drawn
at $z=0.47$, at $t=20,40,60,80$.}
\end{figure}

The role that the lateral boundaries play in the dynamics can be seen by comparing simulations with
Type II and Type IV BCs. The latter involve a square cross-sectioned container with six no-slip boundaries.
The evolution is similar to spin-up in a closed container, with radially outwards flow at the upper and 
lower surfaces (compare Figure \ref{fig:typeIV_sheets} with Figure \ref{fig:schematic_freeslip_vs_noslip}(b)). Because these boundary layers eventually 
reach the lateral surfaces, the container geometry creates alternating sheets of up- and down-welling 
convection that take the form of square annuli. The foregoing argument implies that ring formation with 
the no-slip top surface in the \cite{vorobieff1998} experiments is strongly influenced by the cylindrical shape of the container. 

\subsection{Type V BCs \label{subsec:Type-V-BCs}}

\begin{figure}
\noindent \begin{centering}
\includegraphics[width=0.5\columnwidth]{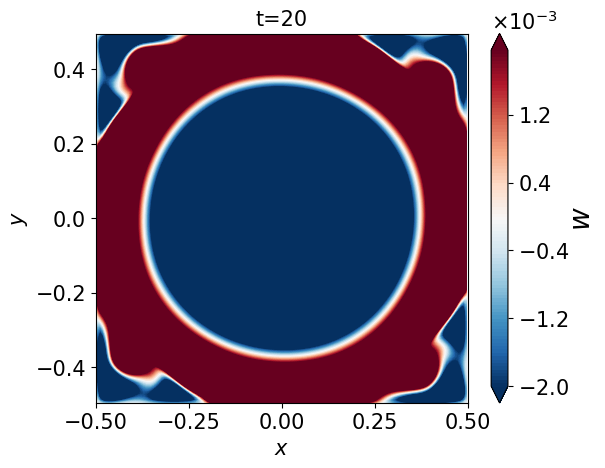}\includegraphics[width=0.5\columnwidth]{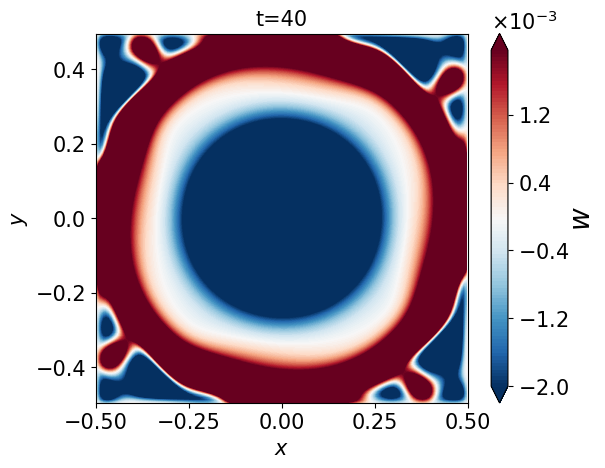}
\par\end{centering}
\noindent \begin{centering}
\includegraphics[width=0.5\columnwidth]{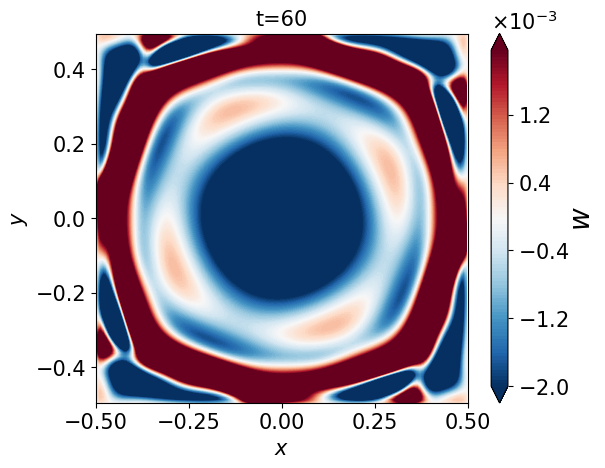}\includegraphics[width=0.5\columnwidth]{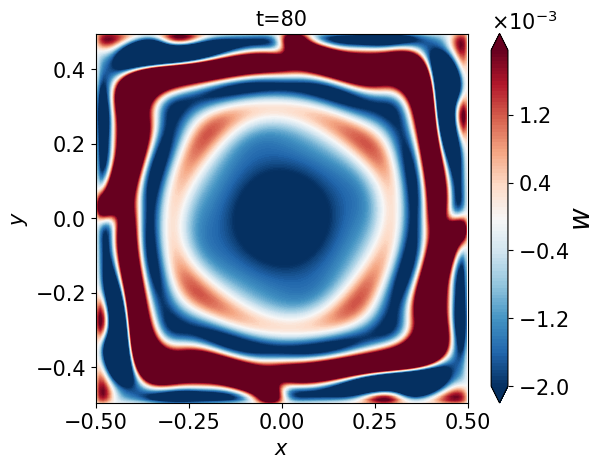}
\par\end{centering}
\caption{\label{fig:typeV_sheets}Evolution of the square sheets of up- and
down-welling in simulations with Type V BCs \tb{for $Re=5000, Ro=0.04, Pr=5$.}. The horizontal sections
are drawn at $z\approx0.025$ at $t=20, 40, 60, 80$ as in Figure \ref{fig:typeIV_sheets}.}
\end{figure}

We examine the processes necessary for ring-formation in terms of the nature of the upper surface boundary conditions.  Namely, if the upper surface is one of free-slip, but lacks buoyancy forcing.  For example, when the no-slip bottom surface provides the buoyancy forcing rings do not form, as can be seen in Figure \ref{fig:typeV_sheets}. This
follows from the mechanism described above; the warm fluid at
the bottom surface is centrifuged outwards and collects at the
upper boundary at the periphery of the container, where it remains, 
taking the shape of the container.

Thus, for containers that are not axisymmetric, the necessary and sufficient condition for convective ring formation
during impulsive spin-up is that the surface providing the buoyancy
forcing be stress-free. This criterion explains the apparent disagreement
between the experiments of \cite{boubnov1986} and \cite{vorobieff1998}.
%(We must presume that \cite{boubnov1986} performed
%experiments in a square container with no-slip surfaces, observed no
%rings, and assumed that the same would be true in a cylindrical container).

\subsection{Type III BCs: the influence of Dirichlet vs Neumann thermal BCs
\label{subsec:Type-III-BCs}}

The thermal BCs play an important role in the dynamics of
convective ring formation. Whereas rings form for both Dirichlet and Neumann thermal BCs, their formation times, locations and lifetimes are
markedly different. In addition, the columnar vortical state is less well defined
with Dirichlet than with Neumann BCs.

Since the temperature difference between the horizontal boundaries is prescribed
instead of the buoyancy flux, we use $\Delta T$ to nondimensionlize Equation \ref{eq:temperature_dim}. Hence, the nondimensionalization of \S \ref{subsec:Nondimensionalisation}
is modified, with  the Rossby number defined as $Ro=g\alpha\Delta T/(\Omega^{2}L)$ {(note that $Ro = {Fr}^{-2}$; see Eq. \ref{eq:Froude} ). 
The definitions of the Reynolds and Prandtl numbers remain unchanged. The Rayleigh number is
\begin{equation}
Ra=\frac{g\alpha\Delta TH^{3}}{\nu\kappa}
%=\frac{g\alpha\Delta T}{\Omega^{2}L}\cdot\frac{\Omega^{2}L^{4}}{\nu^{2}}\cdot\frac{\nu}{\kappa}\cdot\left(\frac{H}{L}\right)^{3}
=\frac{Re^{2}~ Ro~ Pr}{A^{3}},\label{eq:Ra_defn_type3}
\end{equation}
along with the Nusselt number, which may be defined as
\begin{equation}
Nu= \left\langle \frac{\overline{\left(\partial\theta/\partial z\right)}_{z=0}}{A} \right\rangle,
\end{equation}
with, as previously, \tb{$\overline{\cdot}$ denoting the average across a given plane and  $\langle \cdot \rangle$ the time average. The time-averages are taken for $300 < t < 600$, as in Eq. \ref{eq:Nusselt_defn}.}\\

Figure \ref{fig:type_III_rings} shows the ring formation for the
case $Re=7500,Pr=5,$ and $Ro=0.03125$, where the evolution can be compared to that
in Figure \ref{fig:type_I_rings} for Type I BCs ($Re=7500, Pr=5, Ro_f=0.00442$). 
However, the first ring forms earlier and at a larger radius for Type III BCs, the difference being associated with the thermal boundary layers.  Namely, 
for Type III BCs, the thickness of the thermal boundary layer changes significantly with time; fluid from the bottom surface ($\theta=0$) is forced towards the top surface ($\theta=1$) where  the boundary layer grows, eventually
becoming thicker than the corresponding case with Type I BCs.  
In Figure \ref{fig:compare_type_III_typeI}, we see that the overall ring structure has a larger radius with Type III BCs and in Figure \ref{fig:theta_isocontour} the surfaces of constant temperature show that the first ring forms at a larger radius and is thinner for Type III BCs.
Moreover, Figure \ref{fig:buoy_flux_typeIII} shows that the ratio of the maximum buoyancy flux to the long-time average is much larger for Type III BCs than for Type I BCs (Figure \ref{fig:buoy_flux_typeI}).  However, after the rings have broken up into vortices, the thicker thermal boundary layers for 
Type III BCs leads to vortices that gather buoyancy from a broader spatial extent and hence are more diffuse relative to those for Type I BCs (compare Figures \ref{fig:type_III_rings} and \ref{fig:type_I_rings}).   

For geostrophic convection with Type III BCs, the Nusselt number should scale with the Rayleigh number as
\begin{eqnarray}
Nu \propto (Ra / Ra_c)^{3}  \implies   Nu \propto {Ra}^3 {Re}^{-4} \label{eq:Nu_vs_Ra_typeIII} ,
\end{eqnarray}
but this scaling is not seen in Figure \ref{fig:Nu_vs_Ra_typeIII}, as opposed to the collapse shown in Figure \ref{fig:Nu_vs_Ra}(b). \tb{The spread in the curve is due to insufficient averaging, and longer-time averages follow the $Nu \sim (Ra/Ra_c)^{3/4} $ power law. }

We conclude this section by noting that the nature of the global heat transport in non-rotating Rayleigh-B\'{e}nard convection is associated with nature of 
the boundary layer-core interaction, modulated by plumes.  This is heuristically similar to our findings, wherein the nature of the thermal boundary layers
differs for Type I and Type III BCs.

\begin{figure}
\noindent \begin{centering}
\includegraphics[width=0.5\columnwidth]{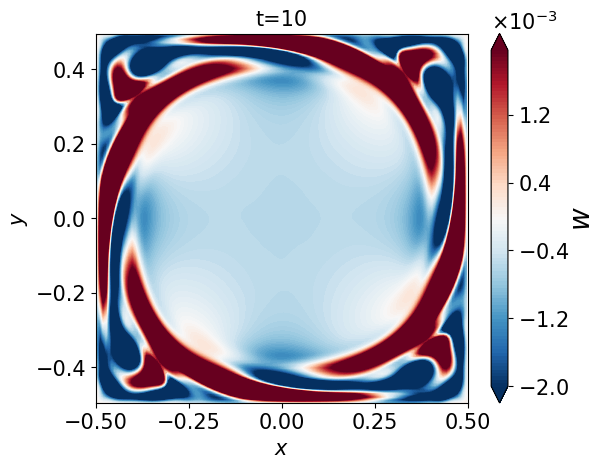}\includegraphics[width=0.5\columnwidth]{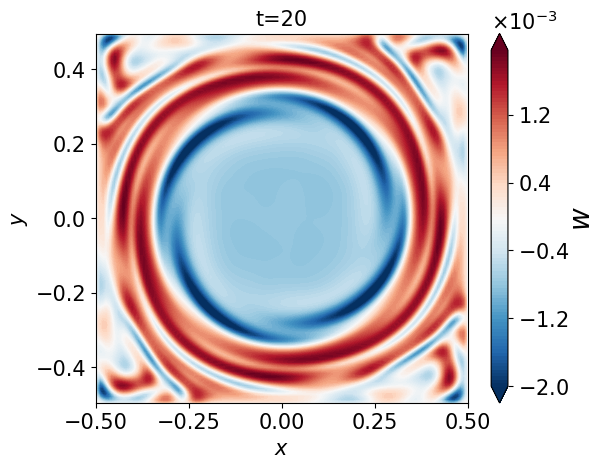}
\par\end{centering}
\noindent \begin{centering}
\includegraphics[width=0.5\columnwidth]{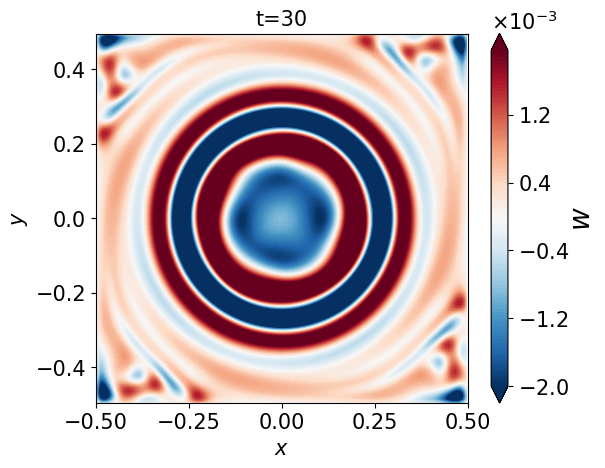}\includegraphics[width=0.5\columnwidth]{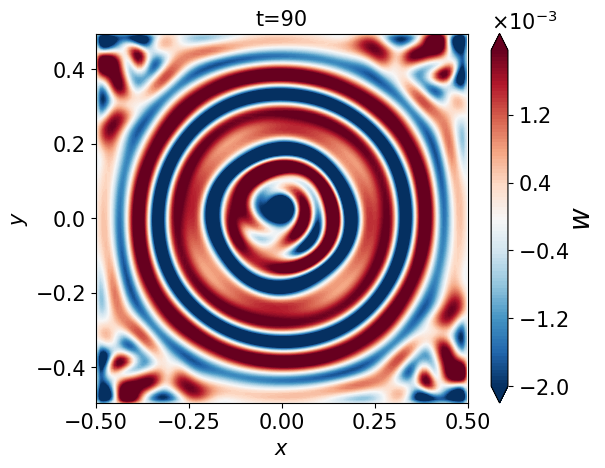}
\par\end{centering}
\noindent \begin{centering}
\includegraphics[width=0.5\columnwidth]{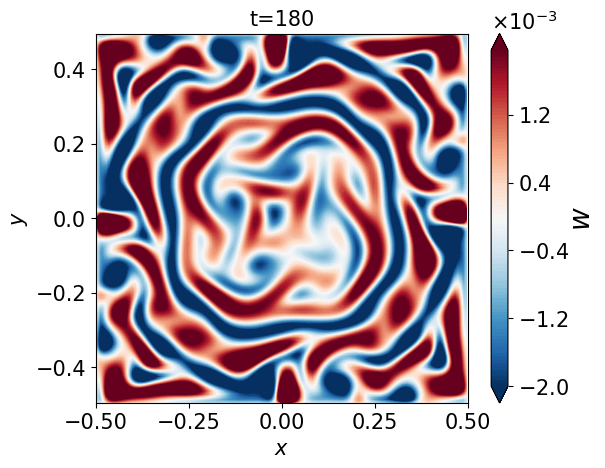}\includegraphics[width=0.5\columnwidth]{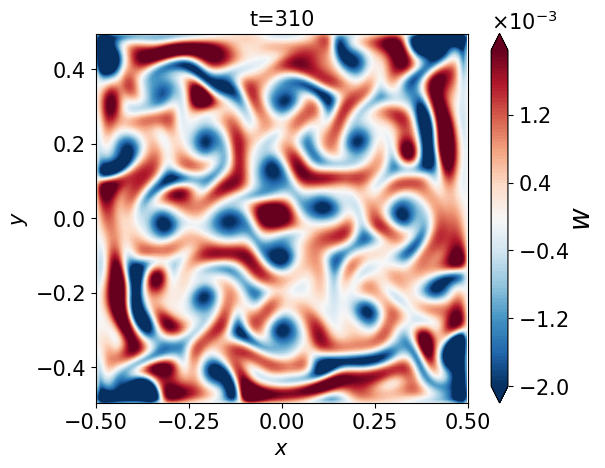}
\par\end{centering}
\caption{\label{fig:type_III_rings} Snapshots of the vertical velocity for Type III BCs at
a horizontal section $z=0.47$ (same as in Figure \ref{fig:type_I_rings}),
for parameters $Re=7500,Pr=5$, (both as in Figure \ref{fig:type_I_rings})
and $Ro=0.03125$.}
\end{figure}

\begin{figure}
\noindent \begin{centering}
\includegraphics[width=1\columnwidth]{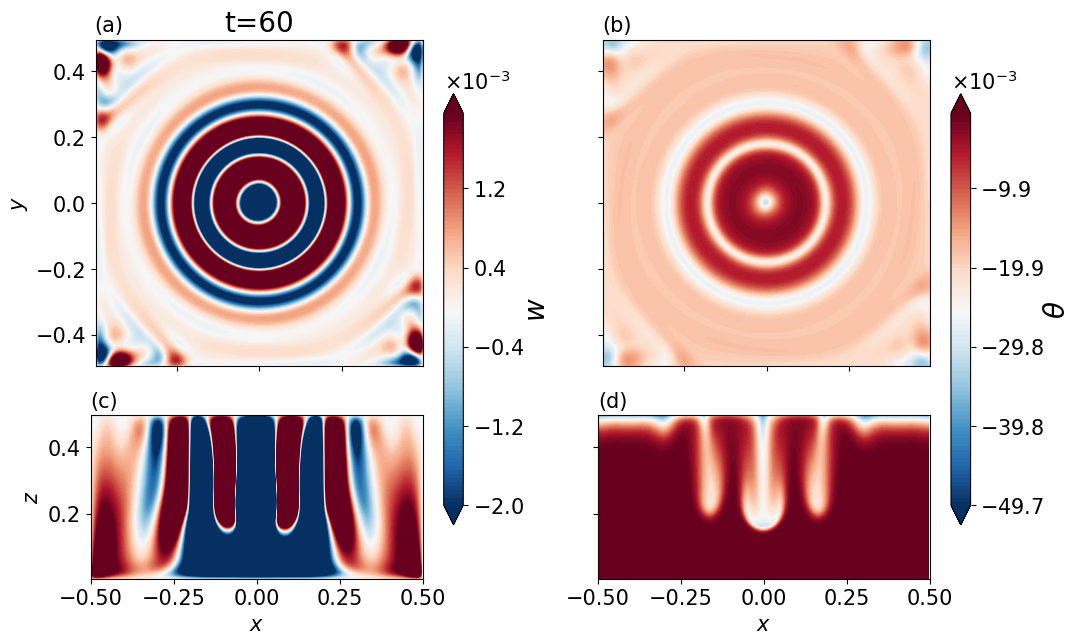}
\par\end{centering}
\noindent \begin{centering}
\includegraphics[width=1\columnwidth]{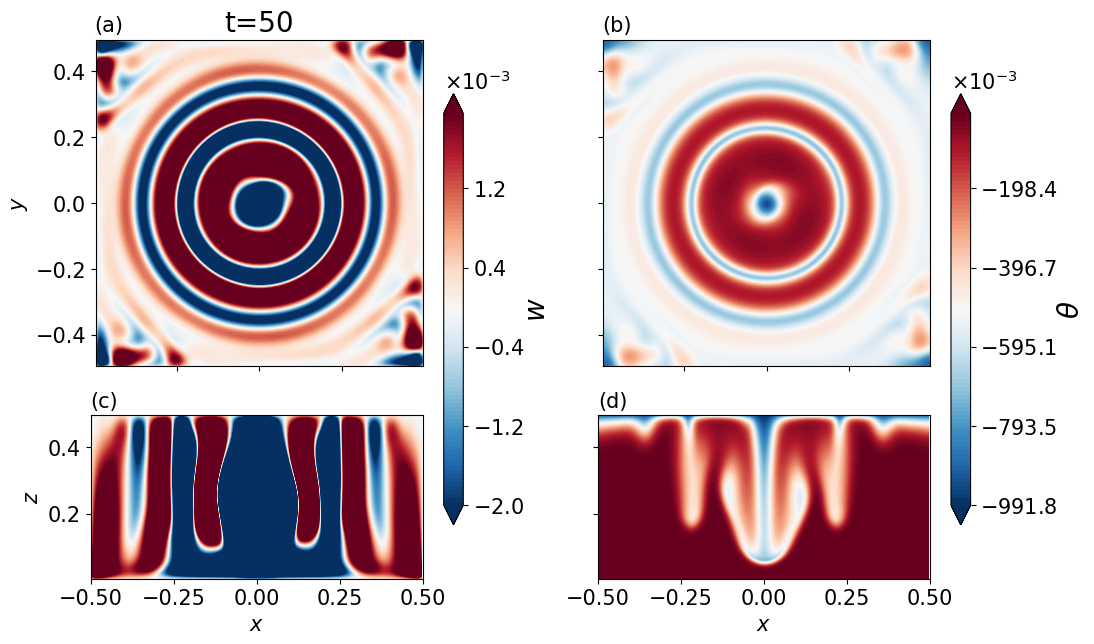}
\par\end{centering}
\caption{\label{fig:compare_type_III_typeI} A comparison of the flow evolution
for Type I and Type III BCs.
%The four figures on top show the state of the flow at $t=60$ for Type I BCs. The four figures in the bottom half show the state of the system at $t=50$ for Type III BCs. 
In each half, the four figures are horizontal sections of the vertical
velocity (a) and temperature (b), and vertical
sections of the vertical velocity (c) and temperature (d). 
The horizontal sections are plotted at $z=0.47$ (same as
in Figure \ref{fig:type_I_rings}). The parameters are $Re=7500,Pr=5$,
(both as in Figure \ref{fig:type_I_rings}) and $Ro_{f}=0.00442$
($t=60$, Type I) and $Ro=0.03125$ ($t=50$, Type III).}
\end{figure}

\begin{figure}
\noindent \begin{centering}
\includegraphics[width=0.5\columnwidth]{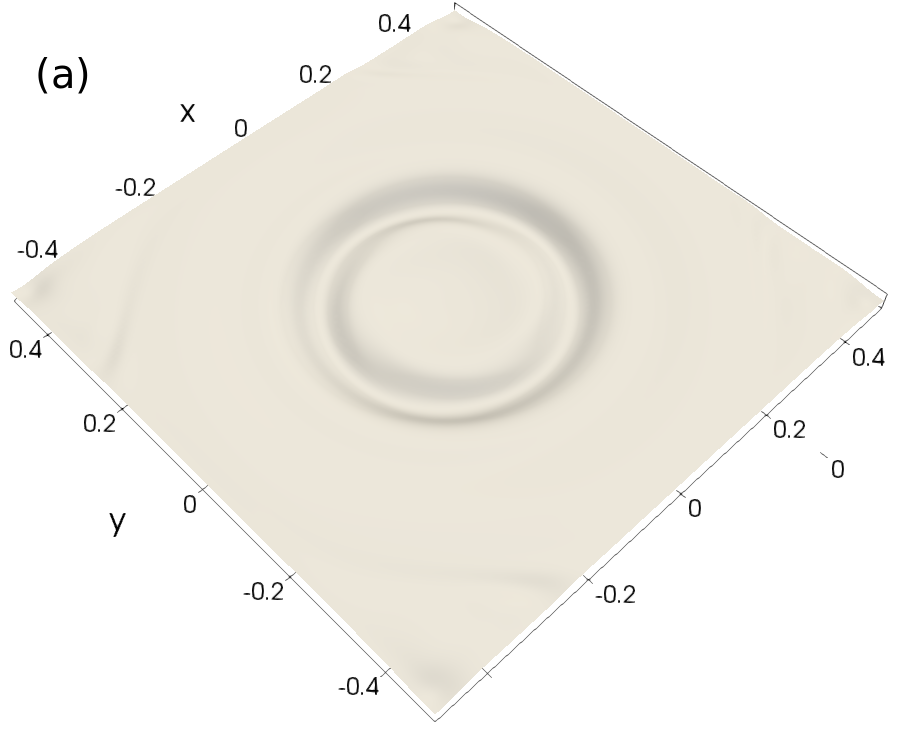}\includegraphics[width=0.5\columnwidth]{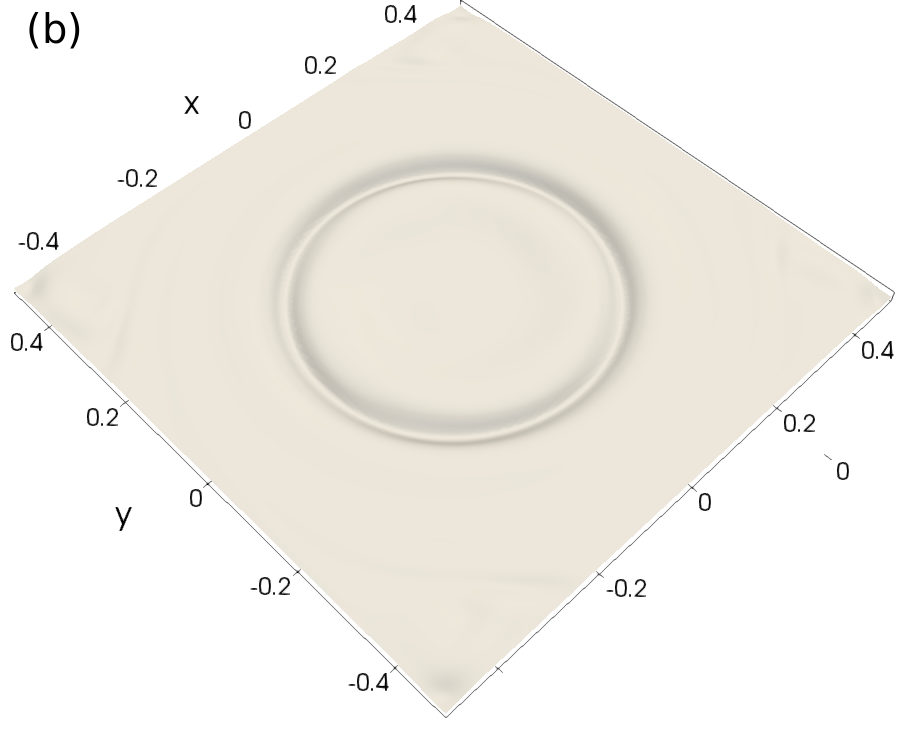}
\par\end{centering}
\caption{\label{fig:theta_isocontour}Isocontours of the temperature, drawn
at a value $\theta^{*}=0.5 \langle \theta \rangle_{\text{free-surface}}$ for (a): Type I 
and $\theta^{*}=0.8 \langle \theta \rangle_{\text{free-surface}}$ (b): Type III BCs. 
The parameters are $Re=7500$, $Pr=5$. $Ro_{f}=0.00442$ ($t=50$)
for the Type I BCs, and $Ro=0.03125$ ($t=30$). for the Type III BCs. These parameters
are the same as in Figures \ref{fig:type_I_rings} and \ref{fig:type_III_rings}.}
\end{figure}

\begin{figure}
\noindent \begin{centering}
\includegraphics[width=1\columnwidth]{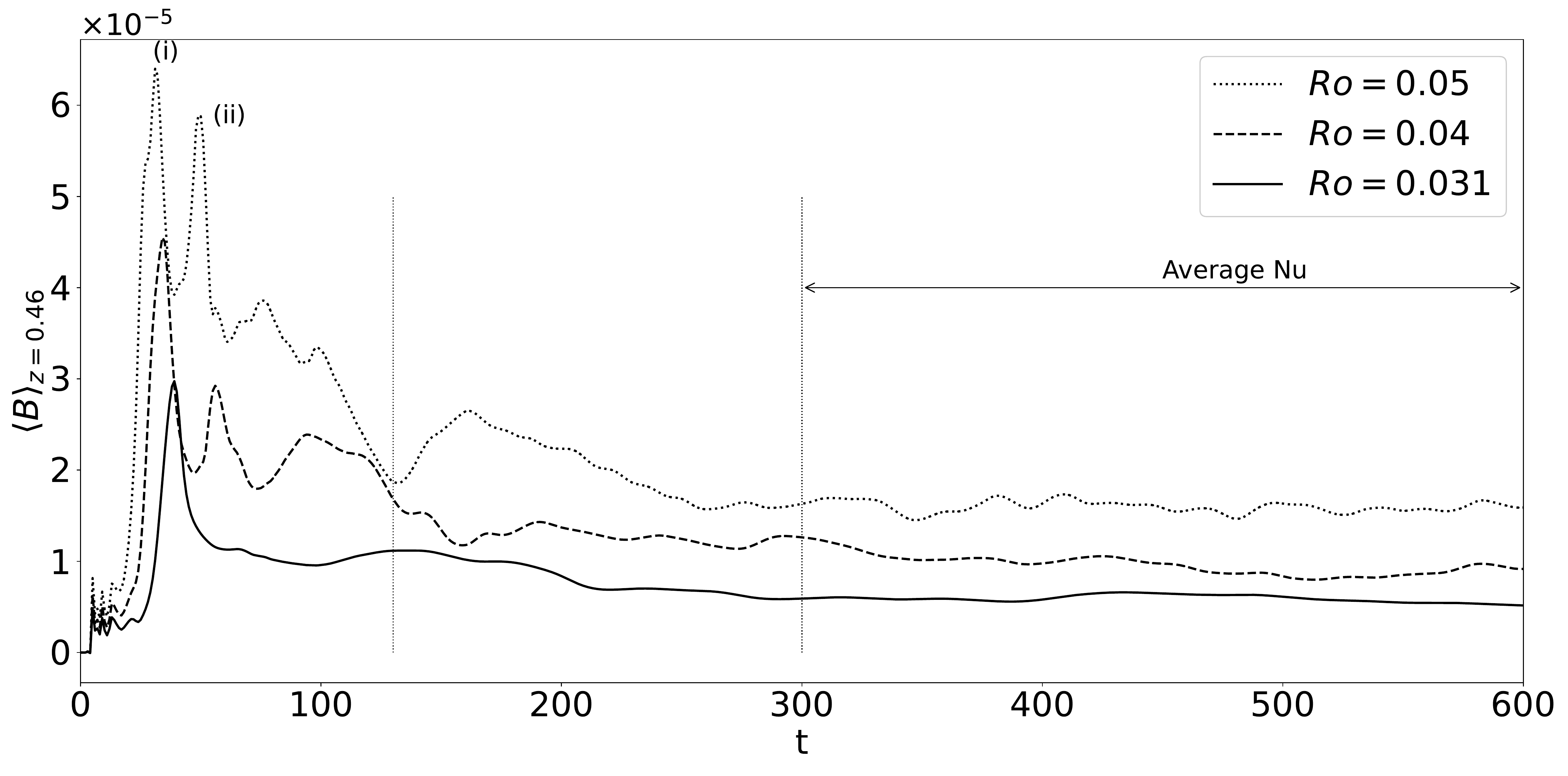}
\par\end{centering}
\caption{\label{fig:buoy_flux_typeIII} The buoyancy flux at $z=0.455$ as
a function of time, for the same parameters ($Re=7500$, $Pr=5$) as in Figure \ref{fig:buoy_flux_typeI},
but with Type III BCs. The peaks of buoyancy flux, labelled (i) and (ii) in the figure,
correspond to the formation of the first ring and the maximal ringed state. 
As with Type I BCs, the time at which the first ring forms decreases for increasing Rossby
number (see Figure \ref{fig:buoy_flux_typeI}). However, the second peaks of the buoyancy flux, corresponding
to the maximally ringed state, occur much sooner here than in Figure \ref{fig:buoy_flux_typeI}.
\tb{As in Fig. \ref{fig:buoy_flux_typeI}, the vertical dotted lines correspond to the completion of spin-up
at $t\approx130$, and the start of averaging for the $\langle Nu \rangle$ calculation respectively.} }
\end{figure}

\begin{figure}
\noindent \begin{centering}
\includegraphics[width=1\columnwidth]{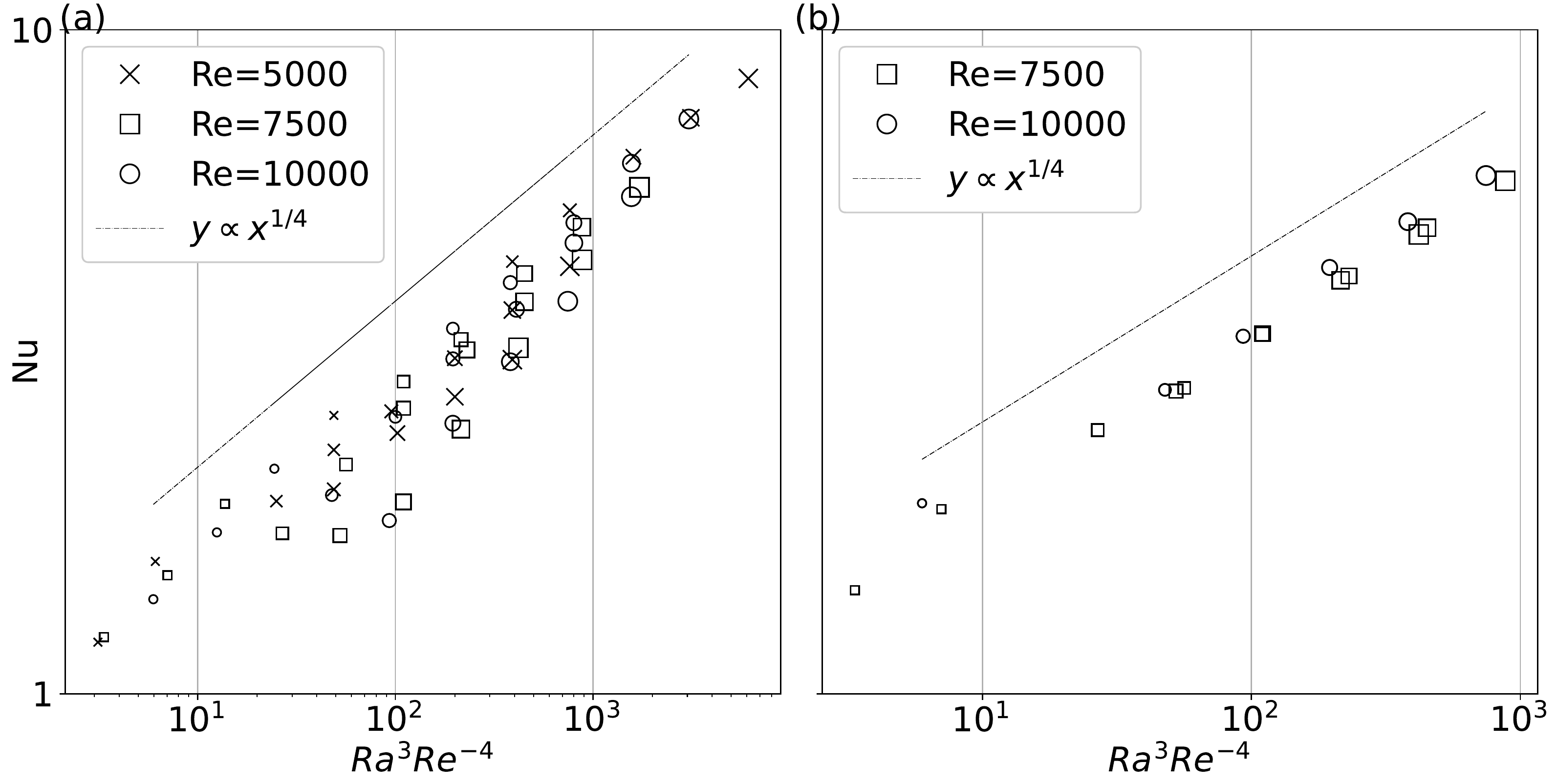}
\par\end{centering}
\caption{\label{fig:Nu_vs_Ra_typeIII} (a) The Nusselt number vs the Rayleigh number
for Type III BCs, analogous to Figure \ref{fig:Nu_vs_Ra} (a). Note that
the Nusselt number does \emph{not} scale like $Ra^{3} Re^{-4}$, but instead
only as $\left(Ra^{3} Re^{-4}\right)^{1/4}$. This lower Nusselt number is responsible for the vortices
being less well defined. \tb{The spread of Nusselt number values for a given Reynolds number is due to insufficient averaging (over $300$ flow units; see Fig. \ref{fig:buoy_flux_typeIII}), and decreases when averaged over longer intervals, as shown in (b), where the Nusselt number is averaged over $3000$ flow units. As in Fig. \ref{fig:Nu_compensated_vs_RaE3by2}, the marker sizes from small to large correspond to $Pr=(1,2,2.5,3.2,4,5)$ respectively, and increase $\propto \sqrt{Pr}$. } }
\end{figure}

\subsection{Special Cases \label{sec:Special_cases}}

As described in \S \ref{subsec:Type-I-BCs}, each step of the
GH spin-up process plays a role in the formation of
convective rings. Thus, altering any of these alters the ring formation process. 
This is seen in the examples presented in \S \ref{subsec:freeslip_side}--\ref{subsec:freeslip_all}
below. Furthermore, a case where the fluid is spun-\emph{down} instead
of spun-\emph{up } is examined in \S \ref{subsec:Convective-Spin-Down}. 

\subsubsection{Free-slip lateral boundaries \label{subsec:freeslip_side}}

The lateral boundaries play an important role in the spin-up process.
GH observe that the diffusion of vorticity from
the lateral surfaces to the fluid results in the suction of flow out
of the boundary layer on the bottom surface into the boundary layers
on the lateral surfaces. It is therefore reasonable to ask; what happens if these are free-slip
surfaces that do not support boundary layers when the no-slip
bottom surface continues to centrifuge fluid outwards?

To this end, Figure \ref{fig:freeslip_sidewalls} shows that while ring formation
does occur, the `rings' are no longer axisymmetric as they were for
the Type I BCs. The radially inward flow in the bulk created
by the boundary layers on the lateral surfaces is thus also responsible for 
pushing the rings that form towards the center, which thereby become
axisymmetric. When these boundary layers are absent, the rings reflect the shape of the container.

\begin{figure}
\noindent \begin{centering}
\includegraphics[width=1\columnwidth]{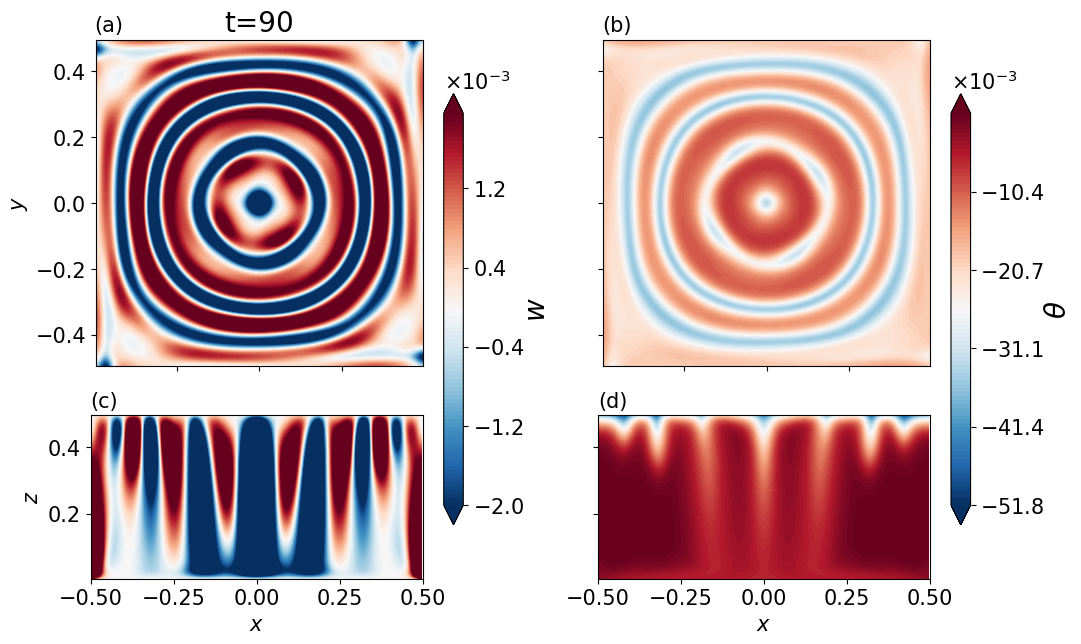}
\par\end{centering}
\caption{\label{fig:freeslip_sidewalls}The formation of convective ``rings''
with free-slip lateral surfaces for the same parameters as in Figure \ref{fig:type_I_rings}.
The horizontal sections of velocity (a) and temperature (b) are plotted at the same 
location $z=0.47$ as in Figure \ref{fig:type_I_rings},
and the vertical sections ((c) and (d) respectively) are plotted on planes
passing through the axes. Note that the bottom boundary is a no-slip
surface.}
\end{figure}

\subsubsection{Free-slip top- and bottom boundaries \label{subsec:freeslip_top_bottom}}

When the top or bottom surfaces obey the no-slip condition, they centrifuge fluid outwards. As we have seen,
this radially outward flow plays a crucial role in the process of ring-formation.
We further illustrate this by making both the top- and bottom-surfaces
free-slip (while the lateral surfaces are no-slip). Rings
form in this case, but at larger radii than in the standard case.
A representative snapshot is shown in Figure \ref{fig:freeslip_top_bottom}.

\begin{figure}
\noindent \begin{centering}
\includegraphics[width=1\columnwidth]{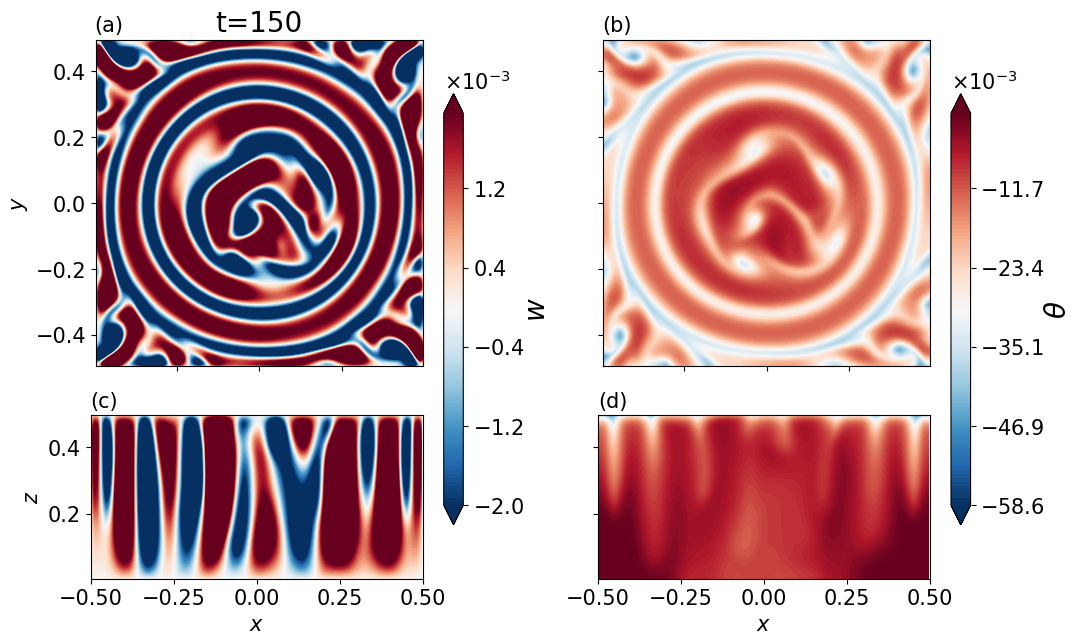}
\par\end{centering}
\caption{\label{fig:freeslip_top_bottom}The formation of convective rings
with free-slip top- and bottom- surfaces for the same parameters as in 
Figures \ref{fig:type_I_rings} and \ref{fig:freeslip_sidewalls}. Note that the lateral boundaries
are here no-slip surfaces. The plots are at the same locations as in, and labelled similarly to, Fig. \ref{fig:freeslip_sidewalls}.}
\end{figure}

\subsubsection{All boundaries free-slip \label{subsec:freeslip_all}}

The examples presented thus far demonsrated the  important role of the boundary
layers on the process of ring-formation. Therefore, it should not be surprising
 that if all the boundaries of the container are made free-slip,
the convective structures that form only have a qualitative resemblance to rings. A representative
snapshot from the evolution of the flow is shown in Figure \ref{fig:freeslip_all},
which should be compared with the evolution in Figures \ref{fig:type_I_rings}
and \ref{fig:freeslip_sidewalls}.

\begin{figure}
\noindent \begin{centering}
\includegraphics[width=1\columnwidth]{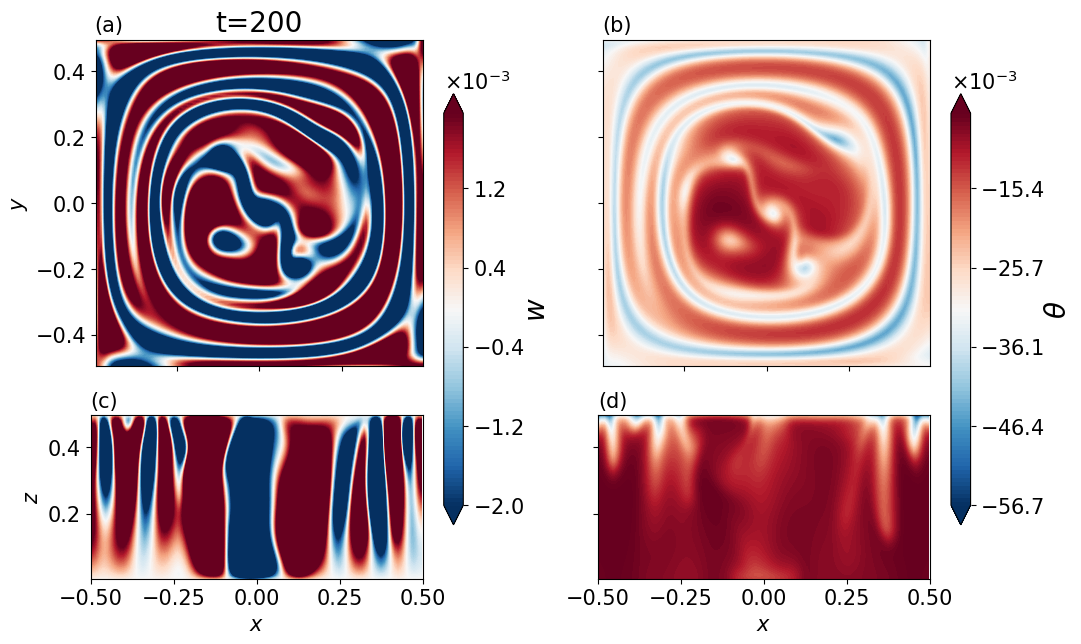}
\par\end{centering}
\caption{\label{fig:freeslip_all} The absence of ring-formation when all the
boundaries are free-slip. The parameters are the same as in Figures
\ref{fig:freeslip_sidewalls} and \ref{fig:freeslip_top_bottom}, and the plots are labelled similarly.}
\end{figure}

\subsubsection{Convective spin-down \label{subsec:Convective-Spin-Down}}

Variations of the mechanism discussed here are also relevant in spin-\emph{down}:
i.e. the case of a rotating container of fluid undergoing a \emph{negative}
step-change in angular speed (at the same moment at which heating/cooling
is switched on at one of its boundaries). In this case the ratio of
initial to final angular velocity (which is zero in spin-up) is also
a parameter. We present here results when the container
slows abruptly from $2\Omega$ to $\Omega$. The fluid velocity at
$t=0$, in the frame of reference rotating with the container, is
thus exactly the negative of the fluid velocity in the spin-up case.
All the other equations remain unchanged. We consider two cases: (a)
with a cooled top surface and (b) with a heated bottom surface. In
both cases, the bottom boundary is no-slip and the upper boundary
is free-slip. 

During spin-down, the flow at the bottom surface is reversed: fluid moves
radially inwards along the surface, and is pushed outwards along the
axis away from the surface. Hence, in case (a) warm fluid impinges
on the top boundary at the axis and moves radially outwards. This
leads to plumes forming at the top surface near the periphery, with
subsequent plumes forming closer to the axis, as shown in Figure \ref{fig:spin_down_top_cooled}.
\tb{Note that there is no boundary layer on the free-slip upper surface. Thus, rings can still form even though the flow is pushed radially outwards.}

\begin{figure}
\includegraphics[width=1\columnwidth]{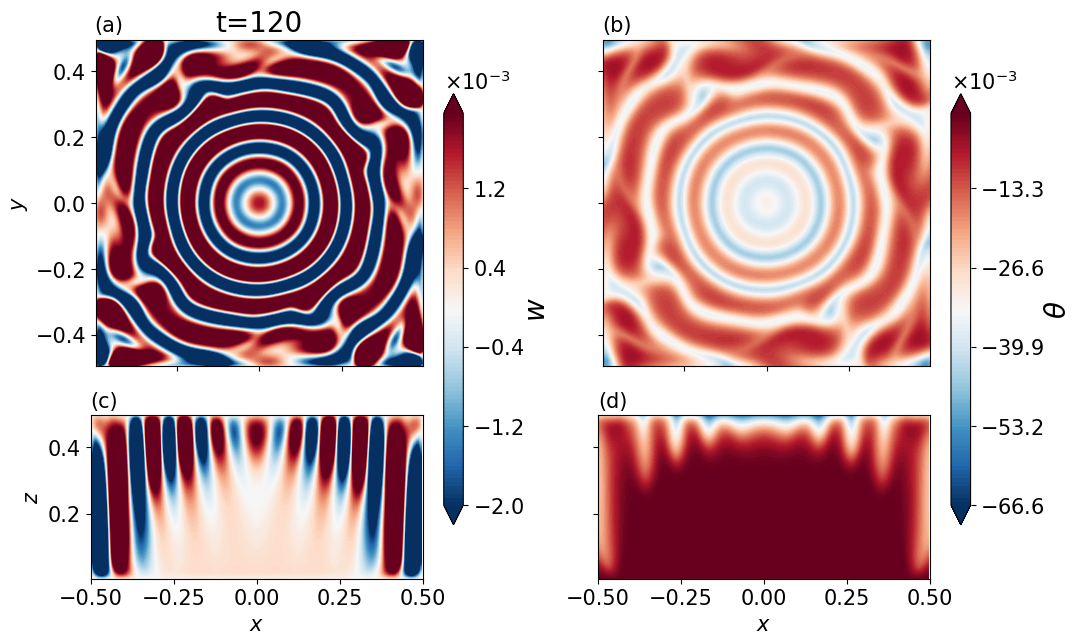}
\caption{\label{fig:spin_down_top_cooled} The formation of rings during top-cooled convective spin-down. The other BCs and parameters are the same as in Figs. \ref{fig:type_I_rings} and \ref{fig:compare_type_III_typeI}, and the figures are plotted in a similar way. The horizontal sections are at $z=0.47$.}
\end{figure}

In case (b), the heating from the surface adds to the Ekman suction at
the bottom surface, with warm fluid forced upwards along the axis. This 
fluid is now at the temperature of the bottom surface, and is pushing
against a background of colder fluid, creating an instability. The interface
splits into rings between the top and bottom surfaces, which break down into vortices as usual. A snapshot of this is shown in Figure \ref{fig:spin_down_bottom_hot}.

\begin{figure}
\includegraphics[width=1\columnwidth]{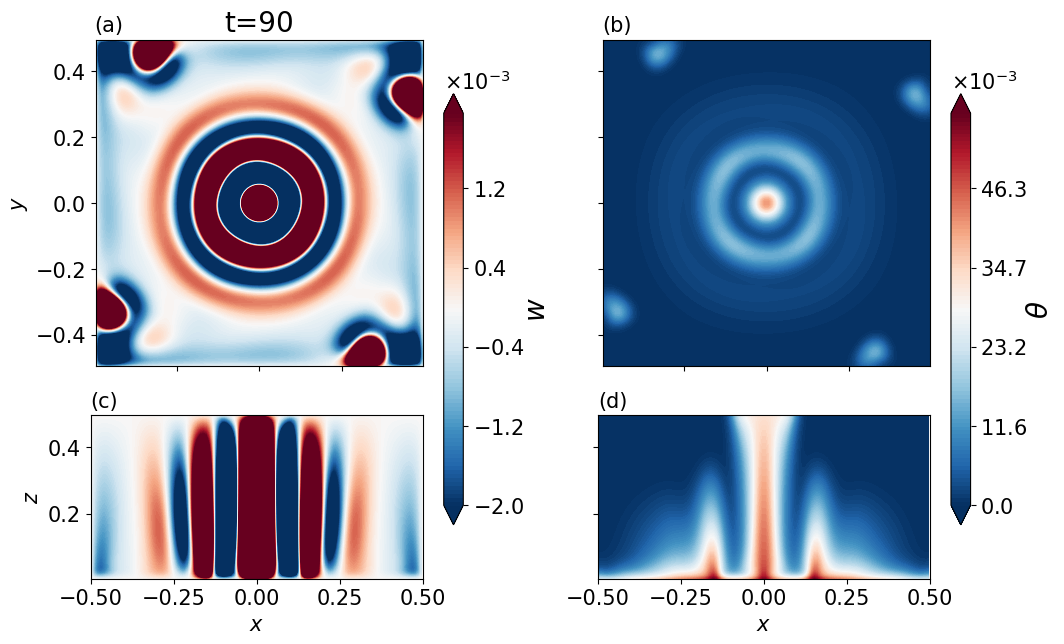}
\caption{\label{fig:spin_down_bottom_hot}The formation of rings during convective spin-down with a heated bottom surface. The other BCs and parameters are the same as in Fig. \ref{fig:spin_down_top_cooled}. The horizontal sections are at $z=0.23$.}
\end{figure}

\section{Conclusion \label{sec:Conclusion}}

In summary, we have performed a range of numerical experiments to
study the formation, longevity and breakdown of a quasi steady ringed
state during the convective spin-up of a Boussinesq fluid. We have
studied the role of the GH spin-up process on ring-formation,
and found that the centrifugal radially outwards flow at the bottom
surface, the reversal of the Ekman layer due to vorticity diffusion at
the side surfaces, and the radially inward flow at the free-slip surface
are all important factors. We show that whereas disrupting {\em any one} of these
disrupts ring formation, and the rings take on the shape of the
container, but disrupting {\em all} of these completely suppresses ring formation.
The ring formation criteria we provide for 
convective spin-up explain the apparent disagreement in experiments
regarding whether rings can form with a solid upper surfaces.

Because the rings arise due to a transient balance between convective and rotational 
dynamics, our finding that the Prandtl number, which we varied from 1 to 5, plays a key role in the formation and stability of the rings is intuitive. We found that the ring lifetime is longest for 
intermediate Prandtl numbers, with a Rossby and Reynolds number dependence.  We have also described the role played by the thermal boundary conditions on the stability of the ringed state and the heat flux in the system. In the transient dynamics considered here, Dirichlet boundary conditions lead to thinner boundary layers and large heat fluxes initially, and lower Nusselt numbers in the steady state, than corresponding cases with Neumann boundary conditions.

\tb{The ring-formation mechanism is general. In results to be reported comprehensively elsewhere, we have observed rings in containers of elliptic cross-section, and in containers of circular cross-section with the lateral walls tapering towards or away from the cooled upper surface. Rings also form in containers with sloped bottom boundaries, and are seen to drift in a direction perpendicular to the slope of the bottom boundary due to a topographic $\beta$ effect. We note that recently \cite{Favier2020wallmodes} have shown that the anticyclonic flow along the edges of a cylindrical container \citep{DeWit2020, Zhang2020} is another geometry-independent universal feature of rotating Rayleigh-B\'enard convection.}

Finally, given the broad relevance of the basic processes we study here, whereby Ekman-layer suction drives the boundary layer fluid 
towards the lateral boundaries at which it may achieve the same speed, a wide range of problems 
may be examined within our general numerical framework through systematic manipulation of the boundary 
conditions to a far greater extent than we have explored here.  Indeed, the generality is extended due to the direct mathematical connection between rotating and stratified fluids \citep{GV:ARFM}, which are uniquely combined in transient rotating convection.   Classical problems 
that arise in this context include those in which  a homogeneous or stratified column of fluid may spin-up or spin-down
due to topographic effects \citep{PDK:DSR} and topographic eddy \citep{HEH:DSR}, Rossby wave \citep{GV:JMR} and edge-wave generation \citep{PBR:Edge}.

%Finally, preliminary results from systems where the boundaries are
%made of a meltable solid show that the dynamics leading to the existence
%of the ringed state seemingly plays a role in more complex systems
%that have no immediate relationship to the spinup problem. The presence
%or absence of columnar vortices controls the Nusselt number, and
%is in turn controlled by the combination of the governing parameters.
%Therefore, in systems where the amount of heat transferred is strongly
%coupled to the dynamics, the dynamics of the system will be strongly
%dependent on the combination of the governing parameters. 

\section*{Acknowledgements}

Computational resources from the Swedish National Infrastructure for
Computing (SNIC) under grants SNIC/2018-3-580 and SNIC/2019-3-386
are gratefully acknowledged. Computations were performed on Tetralith.
The Swedish Research Council under grant no. 638-2013-9243, is 
gratefully acknowledged for support.\\\\
%{\bf Declaration of Interests}. The authors report no conflict of interest.

%
\end{document}